\documentclass[preprint,nofootinbib,amsmath,amssymb]{revtex4-1}

\usepackage{graphicx}
\usepackage{amsmath,color}
\usepackage[normalem]{ulem}

\def\slc#1{\setbox0=\hbox{$#1$}           
    \dimen0=\wd0                                 
    \setbox1=\hbox{/} \dimen1=\wd1               
    \ifdim\dimen0>\dimen1                        
       \rlap{\hbox to \dimen0{\hfil/\hfil}}      
       #1                                        
    \else                                        
       \rlap{\hbox to \dimen1{\hfil$#1$\hfil}}   
       /                                         
    \fi}

\begin{document}

\newcommand{\todo}[1]{(\textbf{\color{red}TODO:} #1)}
\newcommand{\reference}{[\textbf{\color{red}REF}]}
\newcommand{\dd}[2]{\frac{{\rm d}#1}{{\rm d}#2}}
\newcommand{\ud}{\mathrm{d}}
\newcommand{\ie}{{i.e.}}
\newcommand{\rem}[1]{{\color{red} \sout{#1}}}
\newcommand{\newtext}[1]{{\color{green} \bf #1}}

\title{Monoenergetic Gamma-Rays from Non-Minimal Kaluza--Klein Dark Matter Annihilations}

\author{Johan Bonnevier}
\email{bonnev@kth.se}

\author{Henrik Melb\'eus}
\email{melbeus@kth.se}

\author{Alexander Merle}
\email{amerle@kth.se}

\author{Tommy Ohlsson}
\email{tommy@theophys.kth.se}
\affiliation{Department of Theoretical Physics, School of Engineering Sciences,\\
Royal Institute of Technology (KTH),\\
Roslagstullsbacken 21, 106 91 Stockholm, Sweden}

\allowdisplaybreaks

\begin{abstract}

We investigate monoenergetic gamma-ray signatures from annihilations of dark matter comprised of $Z^1$, the first Kaluza--Klein excitation of the $Z$ boson, in a non-minimal Universal Extra Dimensions model. The self-interactions of the non-Abelian $Z^1$ gauge boson give rise to a large number of contributing Feynman diagrams that do not exist for annihilations of the Abelian gauge boson $B^1$, which is the standard Kaluza--Klein dark matter candidate. We find that the annihilation rate is indeed considerably larger for the $Z^1$ than for the $B^1$. Even though relic density calculations indicate that the mass of the $Z^1$ should be larger than the mass of the $B^1$, the predicted monoenergetic gamma fluxes are of the same order of magnitude. We compare our results to existing experimental limits, as well as to future sensitivities, for image air Cherenkov telescopes, and we find that the limits are reached already with a moderately large boost factor. The realistic prospects for detection depend on the experimental energy resolution.

\end{abstract}

\maketitle

\section{Introduction}

There is compelling observational evidence for a dark component of the matter content in the Universe, and yet, the particle nature of this dark matter (DM) remains unknown. Since there is no suitable particle DM candidate in the Standard Model (SM) of particle physics, the need for DM points towards physics beyond the SM. The most popular class of particle DM is weakly interacting massive particles (WIMPs), \ie, particles that are weakly interacting only and have masses in the GeV to TeV range. The standard example of such a DM candidate is the neutralino in supersymmetric extensions of the SM~\cite{Ellis:2010kf}.

WIMP DM candidates also appear naturally in certain extra-dimensional extensions of the SM, \ie, so-called Kaluza--Klein (KK) theories. In particular, in the Universal Extra Dimensions (UED) model, which was proposed by Appelquist, Cheng, and Dobrescu in 2001~\cite{Appelquist:2000nn}, the lightest Kaluza--Klein particle (LKP) is stable, due to conservation of KK parity. In the minimal version of the UED model, the LKP is the first KK excitation $B^1$ of the ${\rm U}(1)_{\rm Y}$ gauge boson $B$~\cite{Cheng:2002iz}, which is non-baryonic and electrically neutral, and hence, a possible DM candidate. This so-called Kaluza--Klein dark matter (KKDM) was first suggested in Refs.~\cite{Servant:2002aq,Cheng:2002ej}. According to detailed calculations of the relic abundance of $B^1$ DM, including the effects of coannihilations, the $B^1$ should have a mass in the range from 500~GeV to 1600~GeV in order to account for the observed DM~\cite{Burnell:2005hm,Kong:2005hn}. Taking resonance effects from second KK level particles into account, it has been shown that the preferred mass range could increase to above 1 TeV \cite{Belanger:2010yx}.

The phenomenology of $B^1$ KKDM has been quite thoroughly investigated in the literature~\cite{Hooper:2007qkref,Arrenberg:2008wy,Belanger:2008gy,Pohl:2008gm,Hooper:2009fj,Flacke:2009eu,Gorchtein:2010xa,Erkoca:2010vk,Bertone:2010fn,Bertone:2010ww,Belanger:2010yx}. However, as shown in Ref.~\cite{Flacke:2008ne}, non-minimal UED models also support other LKPs, which could be DM candidates. Such scenarios give rise to different phenomenological implications, which have only been partially studied in the literature~\cite{Arrenberg:2008wy,Flacke:2009eu,Blennow:2009ag}. 

In this paper, we investigate the annihilation of KKDM into pairs of high-energy photons. This process results in monoenergetic gamma-rays with energy $E_\gamma \simeq m_{\rm LKP}$, which is considered to be a smoking-gun signature for DM. We study a non-minimal UED model, where the main difference to the minimal UED model is that boundary localized terms change the mass spectrum so that the LKP is the $Z^1$. The flux of monoenergetic gamma-rays within the minimal UED model was calculated in Ref.~\cite{Bergstrom:2004nr}. Furthermore, in Ref.~\cite{Bertone:2010fn}, the calculations were repeated, with similar numerical results, and in addition, the contributions to the monoenergetic flux from the final states $\gamma Z$ and $\gamma H$ were calculated. These contributions were concluded to give only a small correction to the total flux. As a check of our calculations, we have successfully reproduced the analytical results in Ref.~\cite{Bergstrom:2004nr}.

Gamma-rays from DM annihilation are currently being searched for by the satellite-borne experiment Fermi-LAT~\cite{Garde:2011wr}, as well as by a number of ground-based image air Cherenkov technique (IACT) telescopes, such as H.E.S.S.~\cite{Abramowski:2011hc}, MAGIC~\cite{Aleksic:2011jx}, VERITAS~\cite{Acciari:2010pja}, and CANGAROO-III~\cite{Kabuki:2007am}. Fermi-LAT is sensitive to gamma-rays with an energy up to approximately 300~GeV, which is too low for the purpose of detection of the monoenergetic gamma-rays studied in this work. On the other hand, IACT telescopes reach energies up to about 10~TeV, and for these searches the limit is instead set by the finite energy-resolution, which may broaden the gamma-ray peak and make it indistinguishable from the continuous background.

The rest of the paper is organized as follows. In Sec.~\ref{sec:TheModel}, we introduce the UED model, including non-minimal versions and implications for the DM phenomenology. Then, in Sec.~\ref{sec:TheProcess}, we describe the analytical calculations for the process $Z^1 Z^1 \to \gamma \gamma$. Next, in Sec.~\ref{sec:Results}, we present our results for the cross section and the flux, and compare them to the case of $B^1$ DM. Finally, in Sec.~\ref{sec:Summary}, we summarize our results and state our conclusions. In addition, in Appendix \ref{sec:FeynmanDiagrams}, we list the Feynman diagrams for the process, and in Appendix~\ref{sec:FeynmanRules}, we present a number of Feynman rules that were used in our calculations.

\section{The UED model}\label{sec:TheModel}

In the UED model, all of the SM fields are promoted to a higher-dimensional flat spacetime, and hence, from the four-dimensional viewpoint, they acquire towers of KK particles. The model is specified by the geometry of the internal space. We consider a single extra dimension with radius $R$, compactified on the flat orbifold $S^1 / {\mathbb Z}_2$, which is the simplest example of a phenomenologically viable UED model~\cite{Cheng:2002iz}. The orbifold is constructed from the circle $S^1$ by identifying points that are connected by the orbifold transformation $y \to -y$. In order for the action to respect the ${\mathbb Z}_2$ symmetry, the five-dimensional fields have to be either even or odd under this transformation, which means that half of the degrees of freedom are removed compared to a full circle $S^1$.

The orbifolding breaks translational invariance along the fifth dimension, and hence, the conservation of extra-dimensional momentum. However, a discrete symmetry known as KK parity, defined as $P_{\rm KK} = (-1)^n$, where $n \in \mathbb{N}$ is the KK number, is conserved. The result is that the LKP is stable, in analogy to the lightest supersymmetric particle in models with $R$ parity conservation.

In a five-dimensional spacetime, the four-component Dirac spinors are irreducible, \ie, there are no chiral spinors. In order for the model to reproduce the SM phenomenology at low-energy scales, the would-be chiral components of the spinors are assigned opposite parity under $y \to -y$, \ie, one is even and the other is odd. Since odd fields have no zero-modes, this assignment results in Dirac spinors that are effectively chiral for energies below $R^{-1}$. At each non-zero KK level, the chiral fermions in the SM are replaced by Dirac spinors with the corresponding quantum numbers.

Compared to the four-dimensional case, a five-dimensional gauge field $A_M$ includes an additional component $A_5$, which behaves as a scalar in the four-dimensional picture. Requiring $A_5$ to be odd under $y \to -y$ removes the massless zero-mode field $A_5^{(0)}$ from the spectrum. At each excited KK level, $A_5^{(n)}$ is eaten by $A_\mu^{(n)}$ to generate the KK mass term, in complete analogy to the SM Higgs mechanism. However, in non-unitary gauges, these fields appear in the loop calculations that are presented in Sec.~\ref{sec:TheProcess}.

In general, the orbifold $S^1 / {\mathbb Z}_2$ allows for boundary-localized terms (BLTs) at the fixed points $y = 0$ and $y = \pi R$. The effect of such terms is to modify the mass spectrum, as well as the coupling constants. In particular, they could affect the identity of the LKP. In the minimal UED model, it is assumed that all BLTs vanish at the cut-off scale of the model, and are only generated at lower-energy scales through renormalization group running~\cite{Georgi:2000ks}. Within this model, the only possible WIMP DM candidate\footnote{However, taking into account the graviton and the full range for the SM Higgs boson mass, the LKP could also be the first KK excitation of the graviton or the charged component of the Higgs field \cite{Cembranos:2006gt}.} is the first KK excitation of the ${\rm U}(1)_{\rm Y}$ gauge boson, the $B^1$. 

In Ref.~\cite{Flacke:2009eu}, it was shown, using a restricted set of BLTs, that it is also possible for the first KK excitations of the $Z$ boson or the Higgs field to be the LKP. In this work, we are interested in the case where the $Z^1$ is the LKP. In general, the BLTs affect the full mass spectrum as well as the coupling constants in a way which is difficult to predict. Here, we make the simplifying assumption that all particles at the first excited KK level, except for the LKP, have equal masses, given by the mass splitting parameter $\sqrt{\eta} = M_1/m_{\rm LKP}$. Using similar assumptions, the authors of Ref.~\cite{Arrenberg:2008wy} performed relic density calculations for the $Z^1$ as the LKP, with the conclusion that its mass should be in the range from 1800~GeV to 2500~GeV.

For the remaining parameter space, we assume that the coupling constants are not affected by the BLTs, which means that the interactions in the non-minimal UED model are identical to those in the minimal one. Then, most of the Feynman rules for the gauge self-interactions can be found in Ref.~\cite{Gustafsson:2008}. However, a few vertices involving $A_5$ fields, as well as the vertices involving Higgs scalars, had to be derived specifically for this work; those are presented in Appendix \ref{sec:FeynmanRules}.

The current experimental bounds on $Z^1$ DM are quite weak. The strongest direct detection limit comes from the XENON100 experiment \cite{Aprile:2011hi}, which constrains the spin-independent DM-nucleon cross section. This quantity was calculated for $Z^1$ DM in Ref.~\cite{Arrenberg:2008wy} for a set of parameter values that are similar to the ones that are used in this work. Comparing the results to the XENON100 limits, we find that, for a relative mass splitting between the $Z^1$ and the first-level KK quarks that is larger than a few percent, the model is only constrained for $Z^1$ masses below approximately 1 TeV, which is too small for the correct relic abundance. In addition, in Ref.~\cite{Blennow:2009ag}, it was found that the indirect neutrino signal from annihilations of $Z^1$ DM particles in the Sun is too weak to be observable in current neutrino telescopes.

\section{The process $\mathbf{\boldsymbol{Z}^1 \boldsymbol{Z}^1 \to \boldsymbol{\gamma} \boldsymbol{\gamma}}$} \label{sec:TheProcess}

As there exist no tree-level Feynman diagrams for the process $Z^1 Z^1 \to \gamma \gamma$, the lowest-order contributions are one-loop diagrams. In our calculations, we take into account only the contributions from the first excited KK level, as higher levels are suppressed by the increasing mass scale. Furthermore, we neglect electroweak symmetry breaking, since the mass scale for the KK modes is much larger than the vacuum expectation value of the Higgs field. This means that all zero-mode particles are effectively taken to be massless.

The main differences between the two-photon annihilation processes for $B^1$ and $Z^1$ DM is that the $Z^1$, being a non-Abelian gauge boson, has self-interactions with the charged ${\rm SU}(2)_{\rm L}$ gauge bosons, and that it only couples to doublet fermions. The latter point has the effect that the contribution from the fermion sector is reduced by a factor of about three. The gauge self-interactions give rise to a large number of relevant Feynman diagrams that do not exist for the analogous $B^1$ process. In order to ensure that all diagrams have been included, we have implemented the model in the LanHEP package \cite{Semenov:2008jy} and used it to generate model files for FeynArts \cite{Hahn:2000kx}, which we in turn have applied to generate the complete set of diagrams for the process. The full set is presented in Appendix~\ref{sec:FeynmanDiagrams}.

For the analytical calculations, we closely follow the method described in Ref.~\cite{Bergstrom:2004nr}. As a check of our calculations, we have been able to exactly reproduce the analytical results of that work for the $B^1$ annihilation process. The amplitude for the process can be written as
\begin{equation}\label{eq:Amplitude}
	\mathcal{M} (p_1,p_2,p_3,p_4) = \epsilon^{\mu}_1(p_1) \epsilon^{\nu}_2(p_2) \epsilon^{\rho}_3(p_3) \epsilon^{\sigma}_4(p_4)\, \mathcal{M}_{\mu\nu\rho\sigma}(p_1,p_2,p_3,p_4),
\end{equation}
where the $\epsilon_i (p_i)$ are the gauge boson polarization vectors. In the approximation that the DM particles annihilate at rest, we have $p_1 = p_2 = p \equiv (m_{\rm LKP}, {\mathbf 0})$, and hence, $2p+p_3+p_4=0$. Using this relation, as well as the transversality of the polarization vectors, the amplitude can be written in the general form
\begin{eqnarray} \label{eq:AmplitudeCoefficients}
	\nonumber \mathcal{M}^{\mu\nu\rho\sigma} (p_1,p_2,p_3,p_4) & = & \phantom{{} + {}} \frac{A}{m_{\rm LKP}^4}~p_3^{\mu}p_4^{\nu}p^{\rho}p^{\sigma} \\
	\nonumber & & {} + \frac{B_1}{m_{\rm LKP}^2}~g^{\mu\nu}p^{\rho}p^{\sigma} + \frac{B_2}{m_{\rm LKP}^2}~g^{\mu\rho}p_4^{\nu}p^{\sigma} + \frac{B_3}{m_{\rm LKP}^2}~g^{\mu\sigma}p_4^{\nu}p^{\rho} \\
	\nonumber & & {} + \frac{B_4}{m_{\rm LKP}^2}~g^{\nu\rho}p_3^{\mu}p^{\sigma} + \frac{B_5}{m_{\rm LKP}^2}~g^{\nu\sigma}p_3^{\mu} p^{\rho} + \frac{B_6}{m_{\rm LKP}^2}~g^{\rho\sigma}p_3^{\mu}p_4^{\nu} \\
	&& {} + C_1~g^{\mu\nu}g^{\rho\sigma} + C_2~g^{\mu\rho}g^{\nu\sigma} + C_3~g^{\mu\sigma}g^{\nu\rho},
\end{eqnarray}
where $A$, $B_i$, and $C_i$ are dimensionless coefficients that depend on the external momenta $p_i$. Equation~\eqref{eq:AmplitudeCoefficients} can be simplified further by using the symmetry for the amplitude of a process with identical bosonic external particles, which gives the following relations for the coefficients
\begin{eqnarray}
	B_2 & = & -B_4,\\
	B_3 & = & -B_5,\\
	B_2 & = & -B_3,\\
	C_2 & = & C_3.
\end{eqnarray}
Indeed, these relations serve as a consistency check of our results.

The gauge invariance of the model has been studied in detail in Refs.~\cite{NovalesSanchez:2010yi,CorderoCid:2011ja}, where it is concluded that the five-dimensional gauge invariance remains in the four-dimensional theory, provided that the KK expansion of the Yang--Mills Lagrangian is performed in the correct way. However, the higher-dimensional theory is a non-renormalizable effective theory, requiring an ultraviolet (UV) completion at some high-energy scale. Hence, the amplitude for the process can generally not be expected to be explicitly finite within the framework of the effective theory. The details of the cancellation of divergences depend on the unknown UV completion, and cannot be explicitly demonstrated without detailed knowledge of this high-energy sector. Thus, we consider only the finite parts of the amplitude, using the fact that the total amplitude has to be finite in any renormalizable theory. For a more detailed discussion on this point, see Appendix~\ref{sec:Divergences}. For the same reason, we cannot use the Ward identity to further simplify the form of the amplitude, as was performed in Ref.~\cite{Bergstrom:2004nr}. 

As a partial check of our implementation of the calculations, we have calculated the amplitude for the similar SM process $Z Z \to \gamma \gamma$ in four dimensions, and confirmed that the Ward identity and the cancellation of divergences hold for that process.

Using the Mathematica package LERG-I~\cite{Stuart:1987tt}, we have calculated the coefficients in Eq.~\eqref{eq:AmplitudeCoefficients}.\footnote{Note that the software LERG-I~\cite{Stuart:1987tt} that we have used is suited only for calculations in the Feynman--'t~Hooft gauge, in which the propagators of the gauge bosons have particularly simple forms. However, we know from Refs.~\cite{NovalesSanchez:2010yi,CorderoCid:2011ja} that our setup is indeed gauge invariant, and all the cross checks we have performed prove the reliability of our results.} Due to the large number of contributing Feynman diagrams, the resulting expressions are complicated and not very illuminating, and hence, we only provide the numerical results presented in Sec.~\ref{sec:Results}. The complete analytical expressions for the contributions from the gauge boson self-interactions can be found in Ref.~\cite{Bonnevier:2011}. Finally, the cross section is given by
\begin{eqnarray}
\sigma v & = & \frac{ \alpha_{\text {SU(2)}}^2\alpha_{{\text {em}}}^2 }{144\pi m_{Z^1}^2} \Big\{ |A|^2+3 |B_1|^2-16|B_2|^2+4 |B_6|^2+12 |C_1|^2+24 |C_2|^2  \nonumber\\
& & + 2 \text{Re}  \left[A \left(B_1^{\ast}+B_6^{\ast}+C_1^{\ast}\right) +B_1 B_6^{\ast} +3 B_1C_1^{\ast}+4 B_6 C_1^{\ast}+2 B_6 C_2^{\ast}+6 C_1 C_2^{\ast} \right] \Big\},
\end{eqnarray}
where we have used the bosonic exchange symmetry to reduce the number of independent coefficients.

\section{Numerical results}\label{sec:Results}

In the numerical calculations, we have employed the same assumptions as in Ref.~\cite{Bergstrom:2004nr}, in order to enable comparisons to the previously existing results. For the coupling constants, we have used the values $\alpha_{\text{em}}^{-1} = 123$ and $\alpha_{{\rm SU}(2)}^{-1} = 95$, which are the values obtained at 1 TeV using the ordinary SM renormalization group running of the physical parameters. As discussed in Appendix~\ref{sec:Divergences}, the accuracy of our predictions depends on the nature of the UV completion of the model, which could, in the worst possible case, give a contribution which is of the same order of magnitude as our results. Hence, unless the UV contribution would almost cancel the contribution that we take into account, our results are valid at least up to an $\mathcal{O} (1)$ numerical factor. Note that such a cancellation would require fine-tuning, which we have no reason to expect.

Let us shortly discuss the continuum part of the spectrum, which has been calculated in Ref.~\cite{Melbeus:2011gs}. In that work, it has been shown that the continuum spectrum is negligible compared to the peak signal investigated here. There are three reasons for this: First, $Z^1 Z^1$ pairs mainly annihilate into pairs of $W$-bosons~\cite{Blennow:2009ag}, which contribute insignificantly to the continuum spectrum near the peak. Second, the number of diagrams contributing to the peak enhances this signal in a significant way, in contrast to the $B^1 B^1$ case~\cite{Bergstrom:2004nr}. Third, since the requirement of having the correct DM abundance forces the mass of the $Z^1$ LKP to be comparatively large, the continuum part of the spectrum is suppressed more strongly compared to the peak for larger DM masses~\cite{Melbeus:2011gs}. Hence, we can safely neglect the continuum spectrum in the following, since it is several orders of magnitude smaller than the peak signal in the relevant energy range.

\begin{figure}[th]
\begin{center}
\includegraphics[width=.6\textwidth]{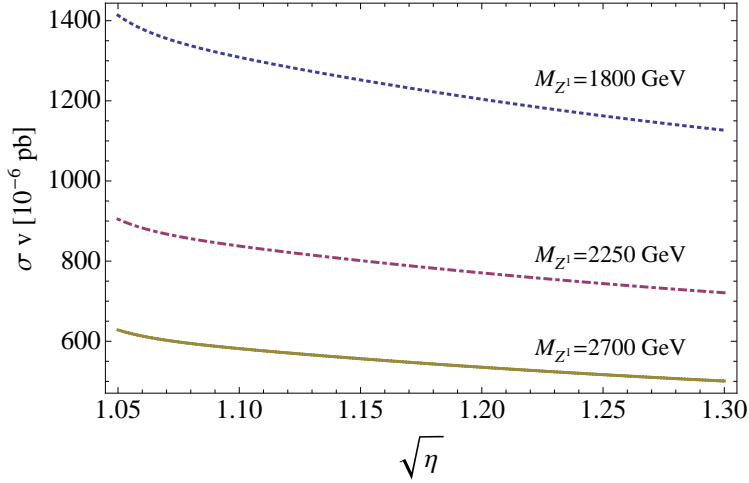}
\caption{The annihilation rate for the process $Z^1 Z^1 \to \gamma \gamma$ as a function of the splitting between the LKP mass and the rest of the first-level excited KK spectrum, $\sqrt{\eta} = M_1 / m_{\rm LKP}$.} \label{fig:csZ1}
\end{center}
\end{figure}

In Fig.~\ref{fig:csZ1}, we show the annihilation rate for the process $Z^1 Z^1 \to \gamma \gamma$ as a function of the mass splitting between the LKP and the rest of the first-level KK spectrum, $\sqrt{\eta} = M_1/m_{\rm LKP}$. The result is plotted for values of the LKP mass of $m_{\rm LKP}$ = 1800~GeV, 2250~GeV, and 2700~GeV, which are the lower end point, the middle point, and the upper end point of the range given by the relic density calculations, respectively. As expected, the rate decreases with increasing mass, and also with increasing mass splitting. The difference in the rates between the highest and the lowest LKP mass is about a factor of two.

\begin{figure}[th]
\begin{center}
\includegraphics[width=.6\textwidth]{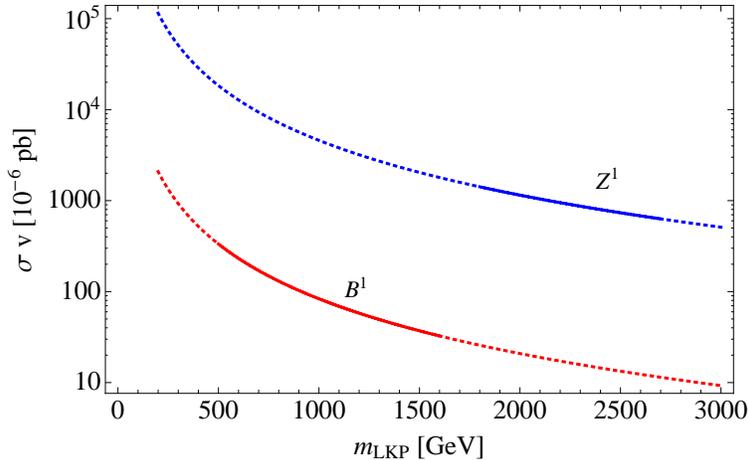}
\end{center}
\caption{The annihilation cross-section for KKDM as a function of the mass, for the mass splitting $M_1 / m_{\text{LKP}} = 1.1$. The solid parts of the curves indicate the mass ranges that give rise to the correct relic abundance for the respective DM candidates.}
\label{fig:Z1_B1}
\end{figure}

The total annihilation rate as a function of the LKP mass, for the $Z^1$ as well as the $B^1$, is shown in Fig.~\ref{fig:Z1_B1}. The solid sections of the curves indicate the mass ranges that reproduce the relic density for the respective LKP candidates, using the results from Ref.~\cite{Arrenberg:2008wy}. Comparing the two options in their relevant mass ranges, it is interesting to note that the rate for $Z^1$ annihilations is larger than the rate for $B^1$ by roughly one order of magnitude, even though $\sigma v \sim 1/m_{\text{LKP}}^2$ and the mass range for $Z^1$ is considerably higher. The reason is simply that, due to the $Z^1$ being a non-Abelian gauge boson, the number of contributing diagrams increases drastically, therefore leading to a significant enhancement of the annihilation cross section. In addition to the $1/m_{\text{LKP}}^2$ behavior, there is a small mass-scale dependence in the $Z^1$ annihilation rate due to the proportionality of some of the vertex factors to $R^{-1}$ (see Appendix~\ref{sec:FeynmanRules}).

Now, the flux at Earth is given by~\cite{Bergstrom:1997fj}
\begin{equation}\label{eq:Flux}
	\Phi (\Delta \Omega) = (2.92 \cdot 10^{-11} \, {\rm m}^{-2} \, {\rm s}^{-1}) \left( \frac{\sigma v}{10^{-29} \, {\rm cm}^3 \, {\rm s}^{-1}} \right) \left( \frac{0.8 \, {\rm TeV}}{m_{\rm LKP}} \right)^2 \langle J_{\rm GC} \rangle_{\Delta \Omega} \ \Delta \Omega,
\end{equation}
where $\Delta \Omega$ is the angular acceptance of the detector and $\langle J_{\rm GC} \rangle_{\Delta \Omega}$ is the averaged line-of-sight integral, which depends only on the DM distribution. We assume a Navarro--Frenk--White (NFW) halo profile~\cite{Navarro:1995iw} with $\alpha = 1.0$, $\beta = 3.0$, $\gamma = 1.0$, and $r_S = 20~{\rm kpc}$. Furthermore, we take $\Delta \Omega = 10^{-5}$, giving $\langle J_{\rm GC} \rangle_{\Delta \Omega} \Delta \Omega = 0.13 b$, where $b$ is a boost factor, which may arise if the DM distribution is clumpy rather than smooth~\cite{Silk:1992bh}. In the experimental data, the peak will be broadened due to the finite energy resolution of any detector. This effect can be represented by a smearing of the differential flux by a Gaussian distribution function. Due to the additional suppression by $1/m_{\rm LKP}^2$ in Eq.~\eqref{eq:Flux}, the flux from $B^1$ particles is of the same order of magnitude as that from $Z^1$ particles, when they are compared for their respective typical masses. Numerically, we obtain the fluxes $\Phi_{Z^1} \simeq 1.2 \cdot 10^{-12} \, {\rm m}^{-2} s^{-1}$ for $m_{Z^1} = 2250~{\rm GeV}$ and $\Phi_{B^1} \simeq 1.1 \cdot 10^{-12} \, {\rm m}^{-2} s^{-1}$ for $m_{B^1} = 800~{\rm GeV}$.

\begin{figure}[th]
\begin{center}
\includegraphics[width=.75\textwidth]{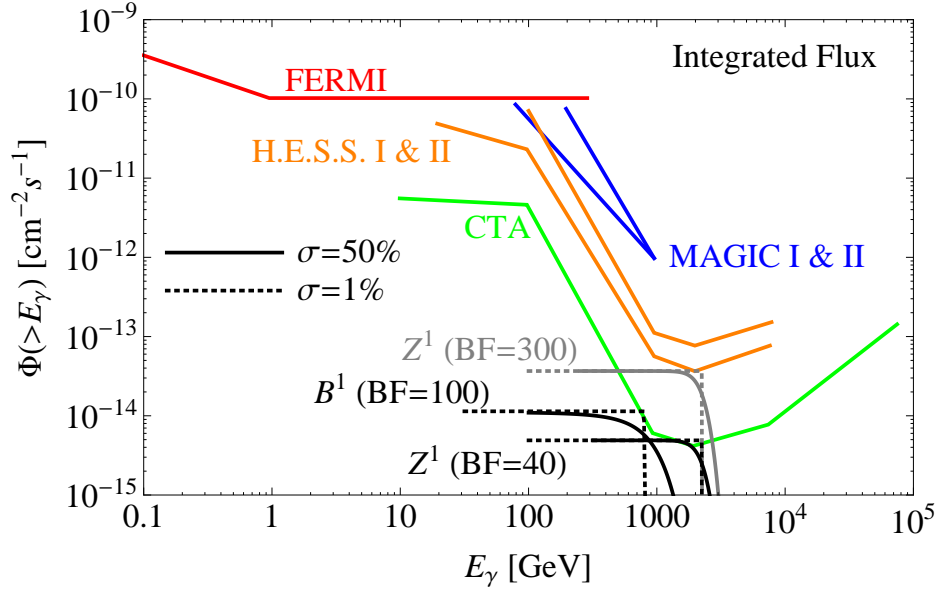}
\caption{The predicted integrated flux above the threshold energy $E_\gamma$, with an experimental energy resolution given by $\sigma$, as a function of $E_\gamma$ for the $Z^1$ and the $B^1$, as well as the upper limits given by a number of experiments. The experimental data is taken from Refs.~\cite{Conrad:2009,Conrad:2011na}.} \label{fig:IntFluxRes}
\end{center}
\end{figure}

Finally, in Fig.~\ref{fig:IntFluxRes}, we show the integrated flux above the threshold energy $E_\gamma$ as a function of $E_\gamma$, and compare the results to existing experimental limits and sensitivities for future experiments. Note that, since the contribution from the continuum spectrum is large only at low-energy scales, the integrated flux above energies close to $m_{\rm LKP}$ are not affected by this contribution. In the mass range that is of interest for this work, the current strongest constraints are given by the H.E.S.S.~collaboration~\cite{Abramowski:2011hc}. Note that the mass range around 1 TeV is where H.E.S.S.~has the best sensitivity. With a boost factor of only $b \simeq 300$, the flux for $Z^1$ DM reaches the limit set by H.E.S.S. In the future, stronger constraints are expected to be set by the Cherenkov Telescope Array (CTA)~\cite{Wagner:2009cs}. These limits are reached already with a boost factor $b \simeq 40$ for the $Z^1$, whereas $b \simeq 100$ would be needed for the $B^1$.

\section{Summary and conclusions}\label{sec:Summary}

We have investigated the annihilation of $Z^1$ DM into monoenergetic gamma-rays, and in addition, we have confirmed the results for the corresponding process for $B^1$ DM that already exists in the literature. Although the $B^1$ is the most frequently studied DM candidate in the context of KKDM, and the one that is obtained in the minimal UED model, the $Z^1$ has been shown to be a viable DM candidate in non-minimal UED models.

Our results illustrate that, even when taking into account the expected larger mass of the $Z^1$ compared to the $B^1$, the annihilation rate for $Z^1$ particles is about one order of magnitude larger than the corresponding rate for $B^1$ particles. However, the flux at Earth, which is further suppressed by the square of the mass, is larger for the $B^1$, although only by a small factor.

The experimental signature of the process would be a peak in the differential flux on top of the continuous gamma-ray spectrum from DM annihilations, which extends up to the energy scale $m_{\rm LKP}$. The prospects for detection depend on the size of the monoenergetic peak in relation to the continuum flux. As was shown in Ref.~\cite{Bergstrom:2004nr}, the characteristic peak could be effectively erased if it is too low and/or the experimental resolution is not good enough.

It is clear from relic abundance calculations that an experiment reaching up to several TeV in energy would be needed to detect the peak. We have shown that the contribution from only the monoenergetic flux would reach the existing H.E.S.S.~limits or upcoming CTA limits already with a moderately large boost factor for $Z^1$ DM, whereas a somewhat larger boost factor would be needed for $B^1$ DM.

\begin{acknowledgments}

We would like to thank Michael Gustafsson and Torsten Bringmann for useful discussions and Jan Conrad for providing useful information.

This work was supported by the Swedish Research Council (Vetenskapsr{\aa}det), contract no.~621-2008-4210 (T.O.), by the G{\"o}ran Gustafsson foundation (A.M.), and by the Royal Institute of Technology (KTH), project no.~SII-56510 (A.M.).

\end{acknowledgments}

\appendix

\newpage

\section{Feynman diagrams}\label{sec:FeynmanDiagrams}

\begin{figure}[htb]
\includegraphics[width=.17\textwidth]{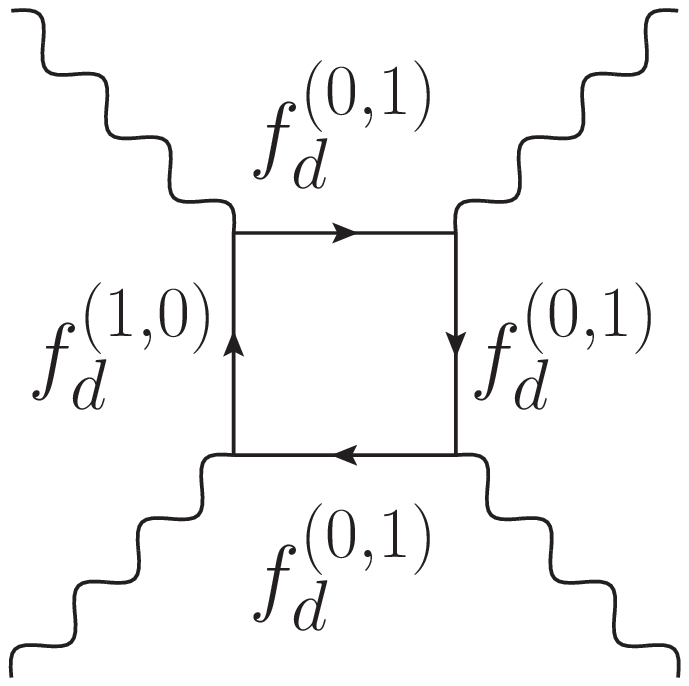} \hspace{.02\textwidth}		
\includegraphics[width=.17\textwidth]{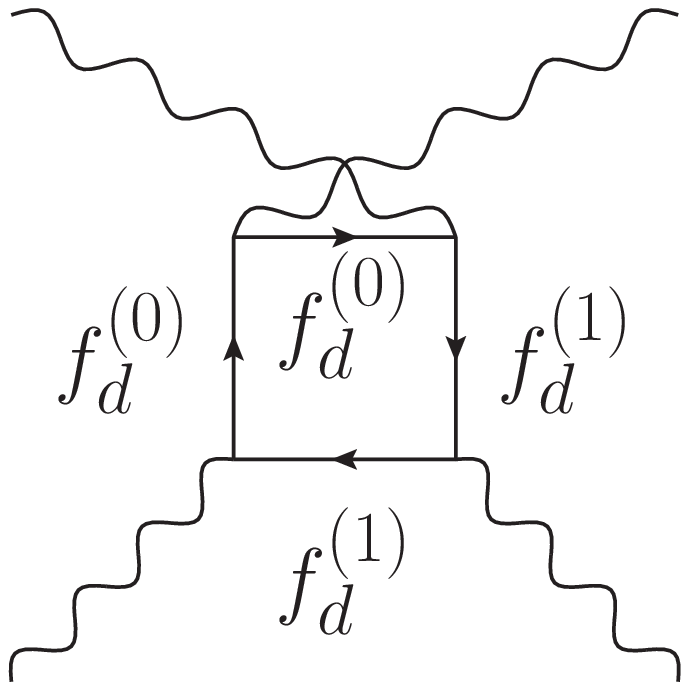}								
\caption{The one-loop level Feynman diagrams involving internal fermion fields for the process $Z^1 Z^1 \to \gamma \gamma$. For each diagram, there is also a corresponding one with crossed final states, making the total number equal to 6. The notation $A^{(0,1)}$ simply indicates that there are two different choices for the flow of KK number in the diagram, and the first or second choice should be used throughout the diagram.} \label{fig:FeynmanDiagramsFermions}
\end{figure}

\begin{figure}[htb]
\includegraphics[width=.17\textwidth]{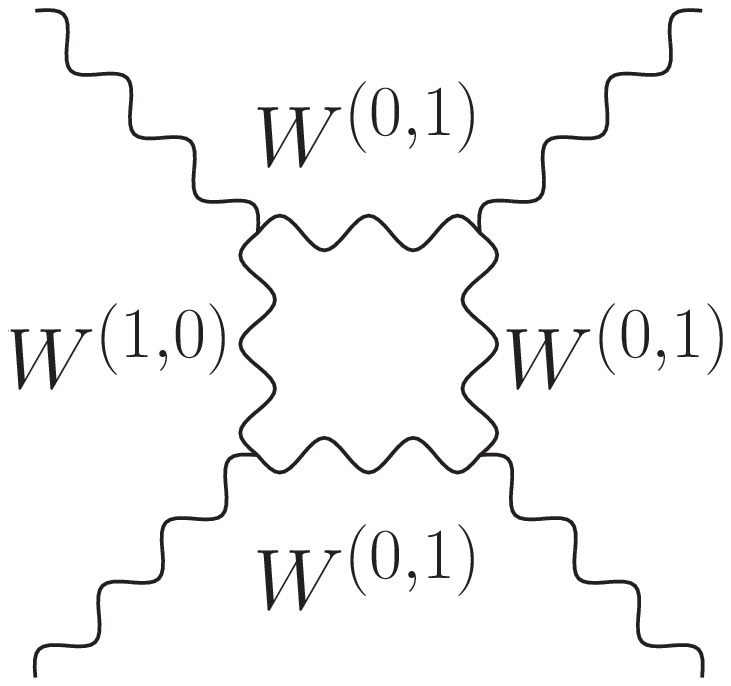} \hspace{.02\textwidth}	
\includegraphics[width=.17\textwidth]{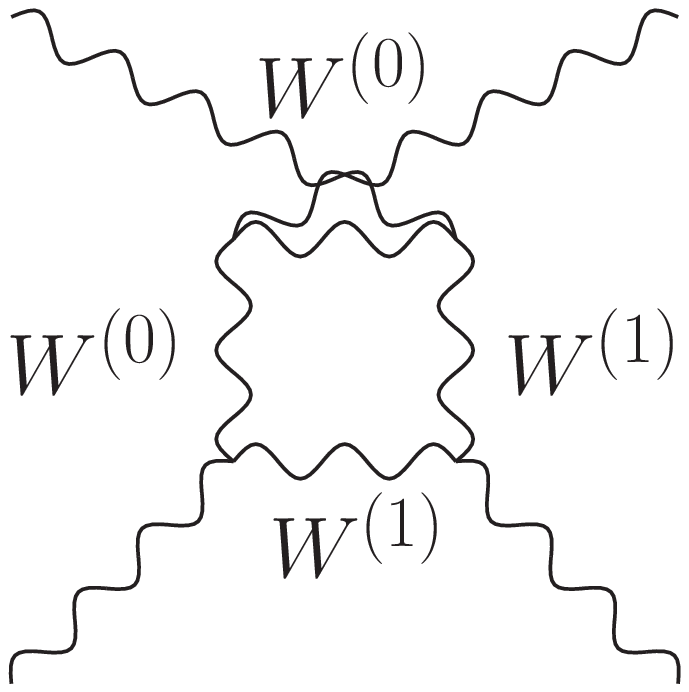} \hspace{.02\textwidth}		
\includegraphics[width=.17\textwidth]{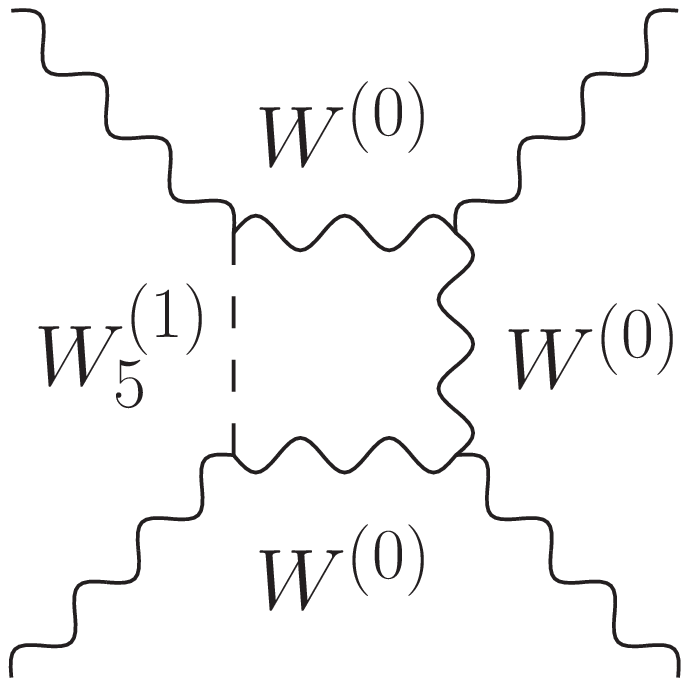} \hspace{.02\textwidth}		
\includegraphics[width=.17\textwidth]{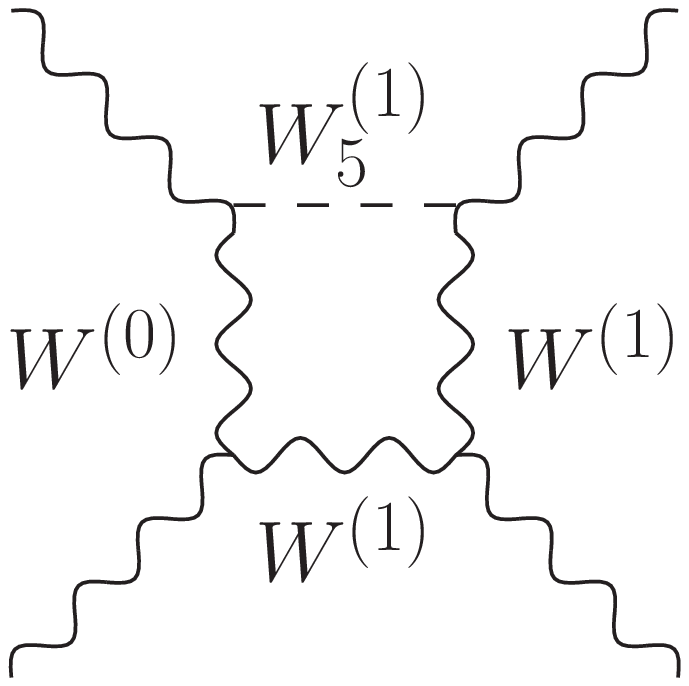} \hspace{.02\textwidth}		
\includegraphics[width=.17\textwidth]{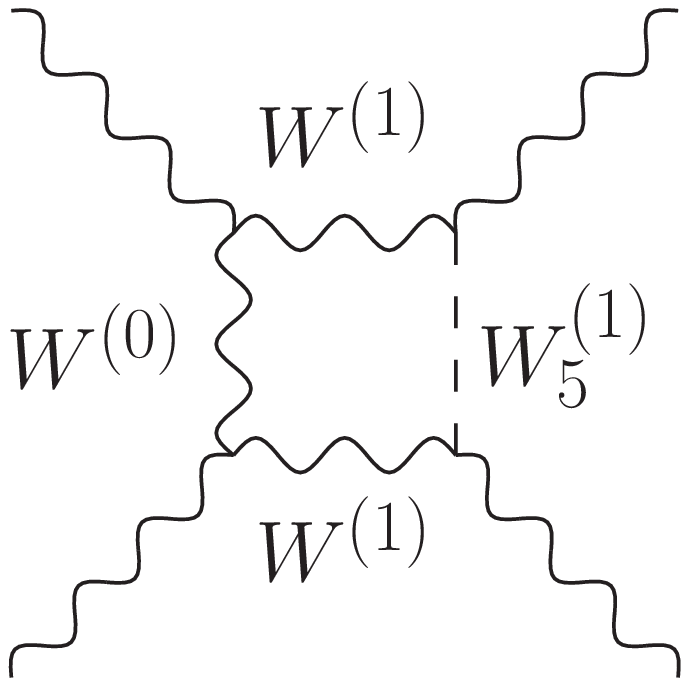} \\							
\vspace{5mm}
\includegraphics[width=.17\textwidth]{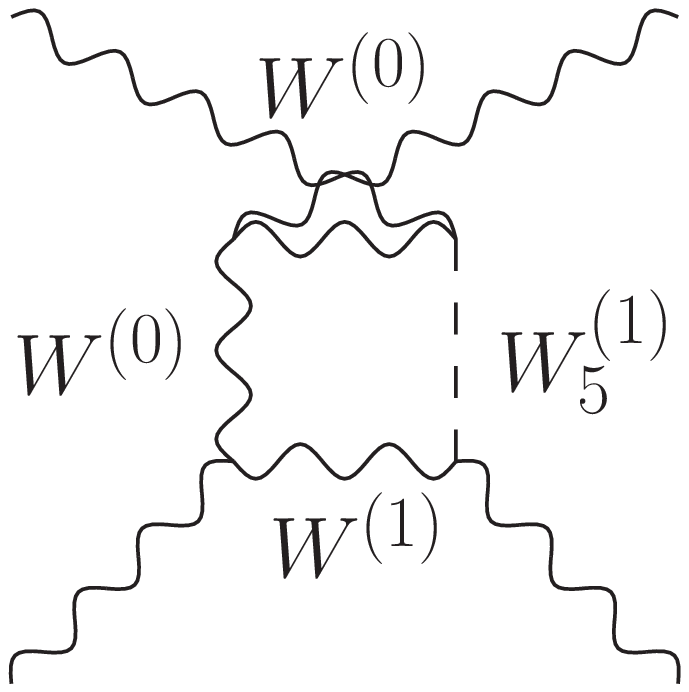} \hspace{.02\textwidth}		
\includegraphics[width=.17\textwidth]{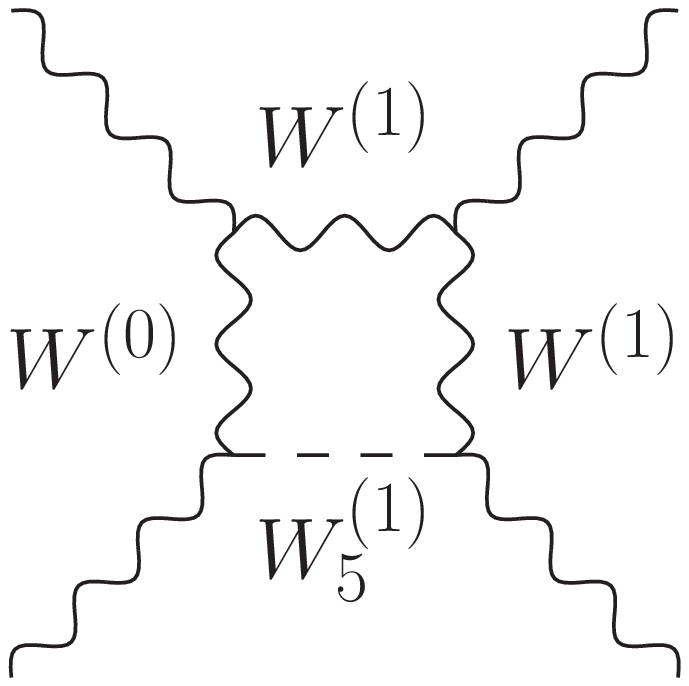} \hspace{.02\textwidth}		
\includegraphics[width=.17\textwidth]{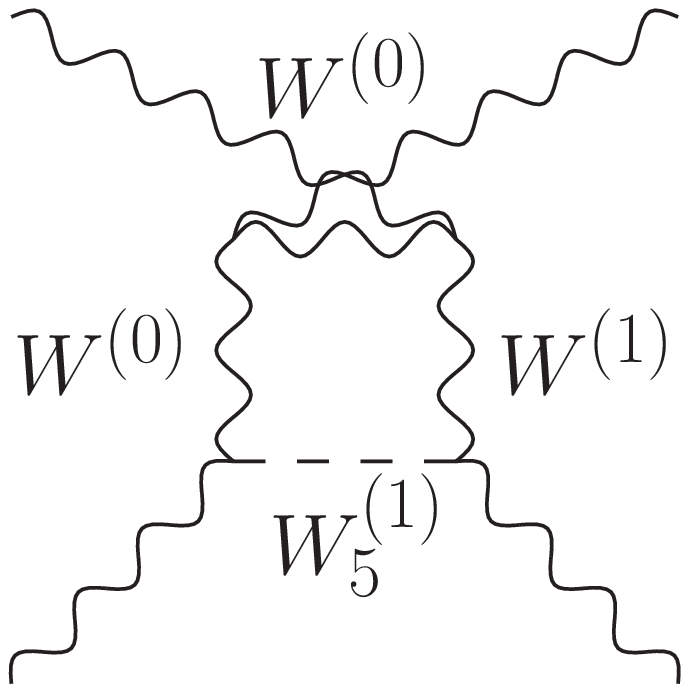} \hspace{.02\textwidth}		
\includegraphics[width=.17\textwidth]{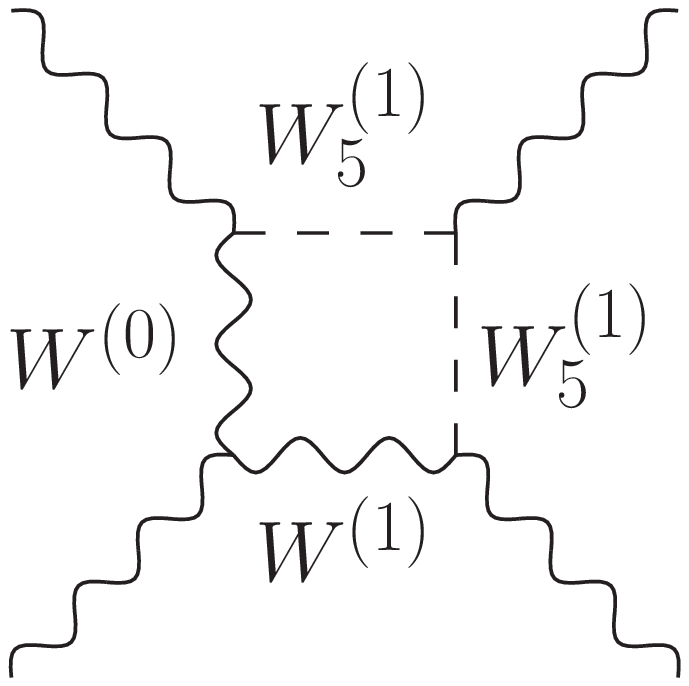} \hspace{.02\textwidth}		
\includegraphics[width=.17\textwidth]{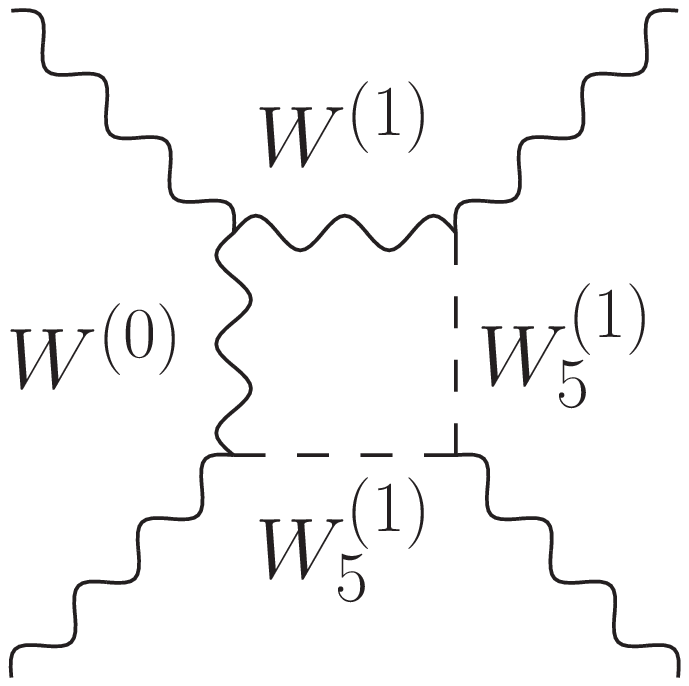} \\							
\vspace{5mm}
\includegraphics[width=.17\textwidth]{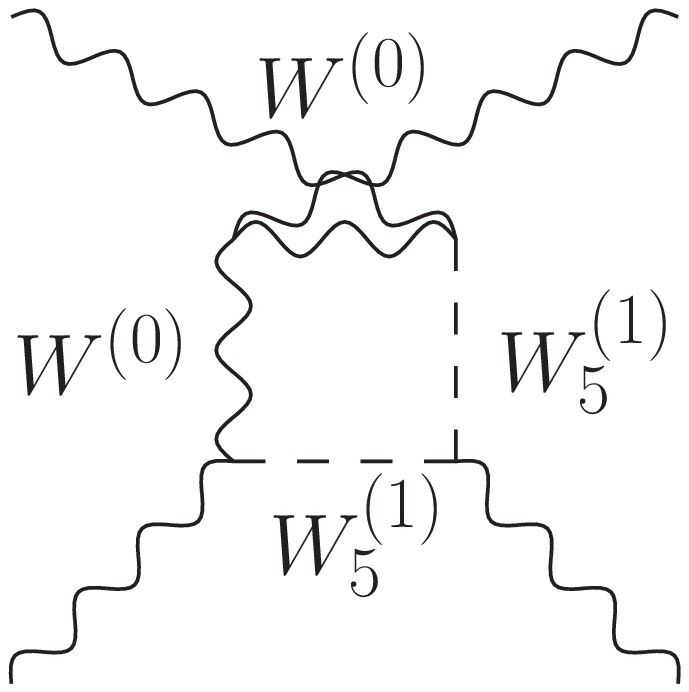} \hspace{.02\textwidth}		
\includegraphics[width=.17\textwidth]{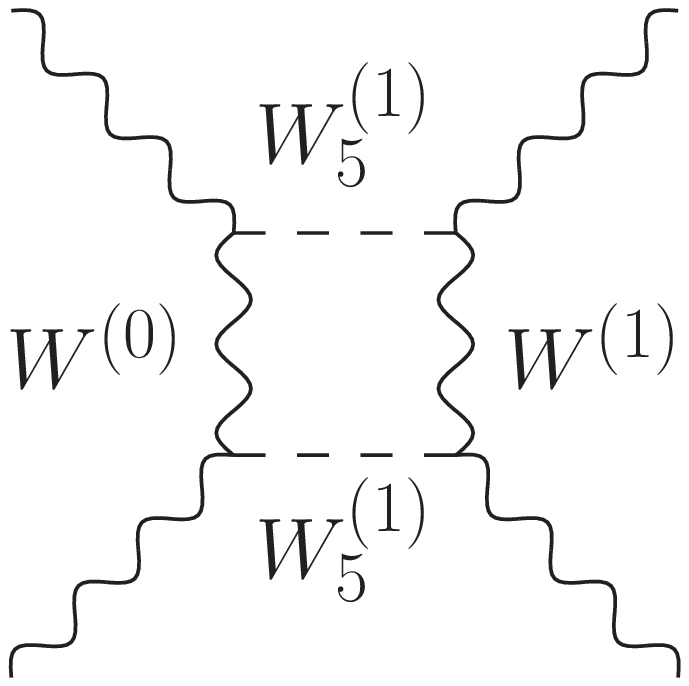} \hspace{.02\textwidth}		
\includegraphics[width=.17\textwidth]{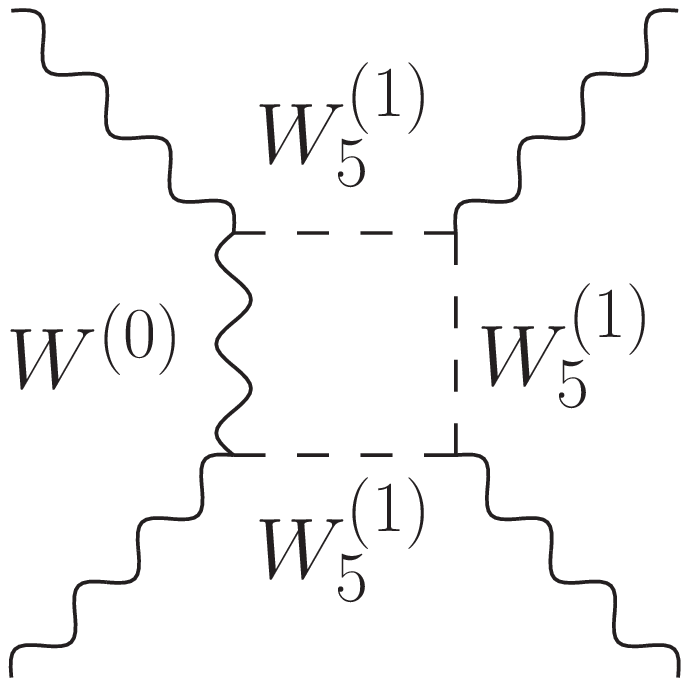} \hspace{.02\textwidth}		
\includegraphics[width=.17\textwidth]{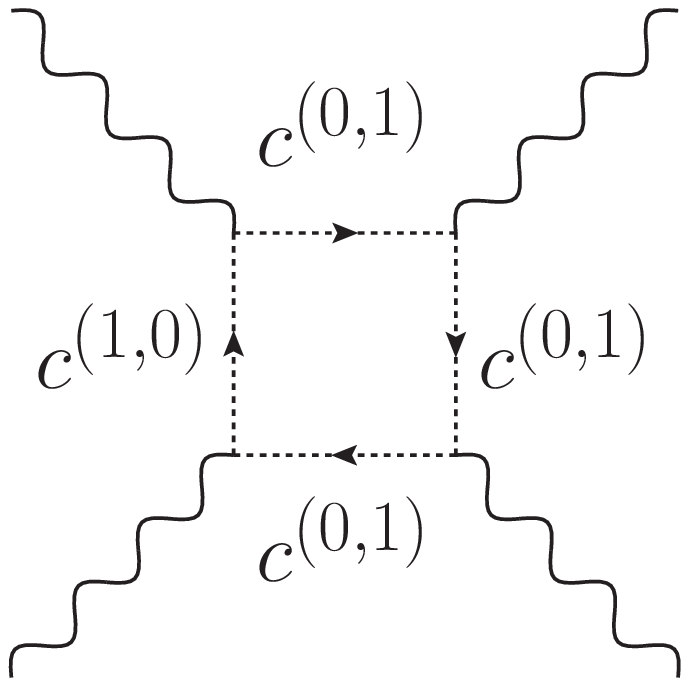} \hspace{.02\textwidth}	
\includegraphics[width=.17\textwidth]{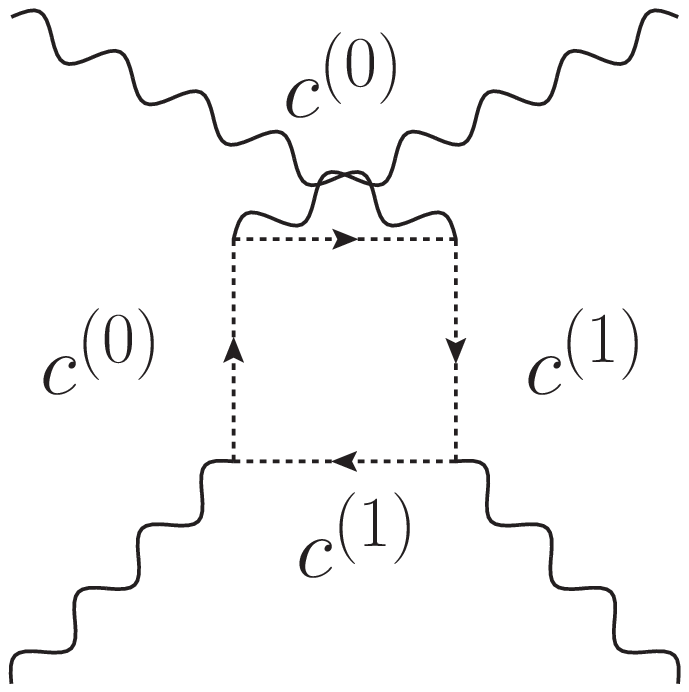} \\							
\vspace{5mm}
\includegraphics[width=.17\textwidth]{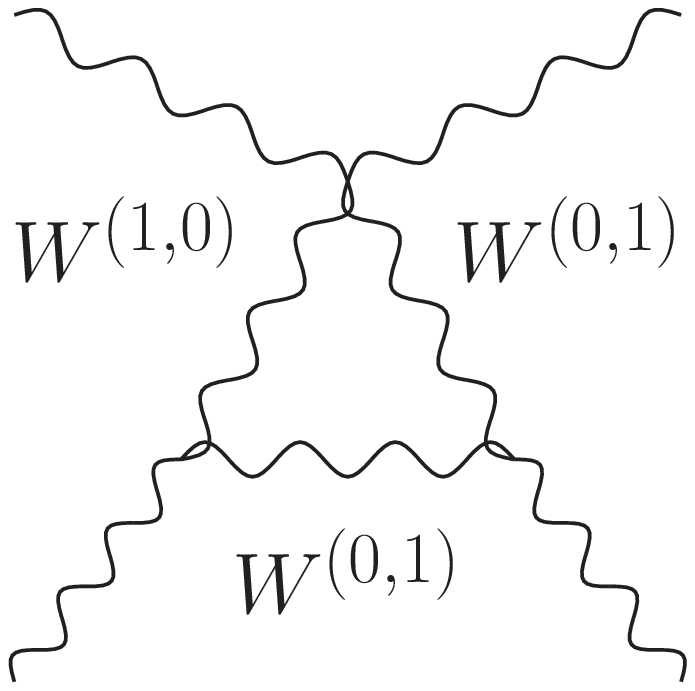} \hspace{.02\textwidth}	
\includegraphics[width=.17\textwidth]{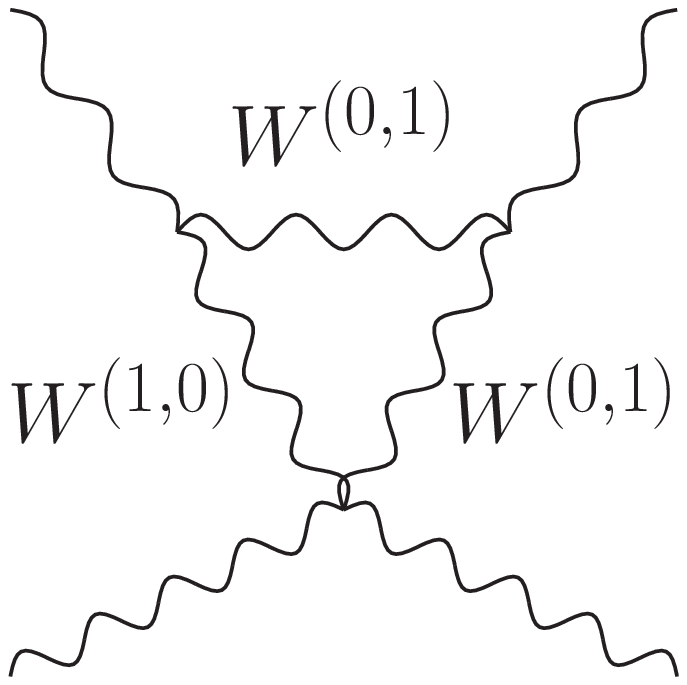} \hspace{.02\textwidth}	
\includegraphics[width=.17\textwidth]{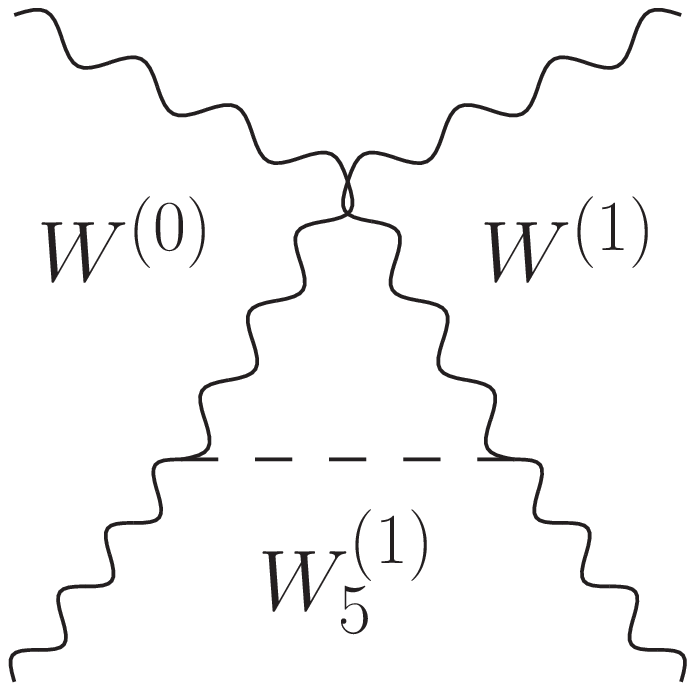} \hspace{.02\textwidth}		
\includegraphics[width=.17\textwidth]{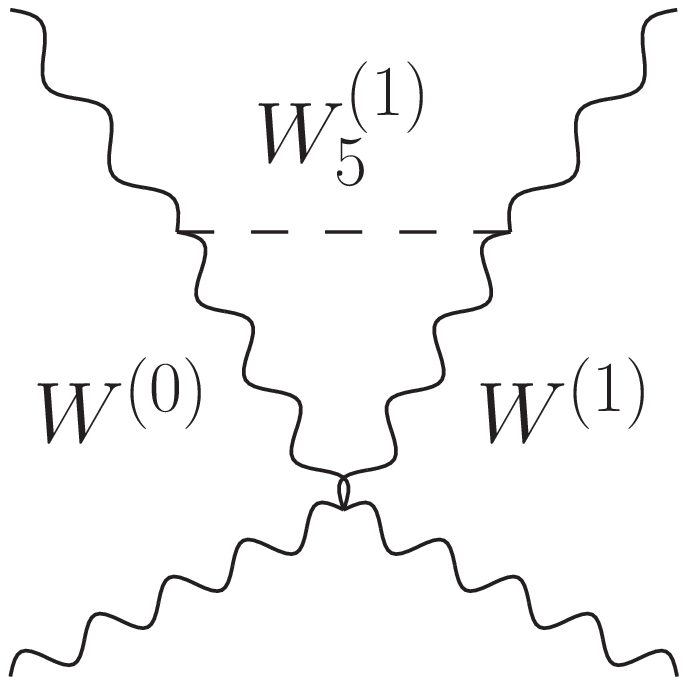} \hspace{.02\textwidth}		
\includegraphics[width=.17\textwidth]{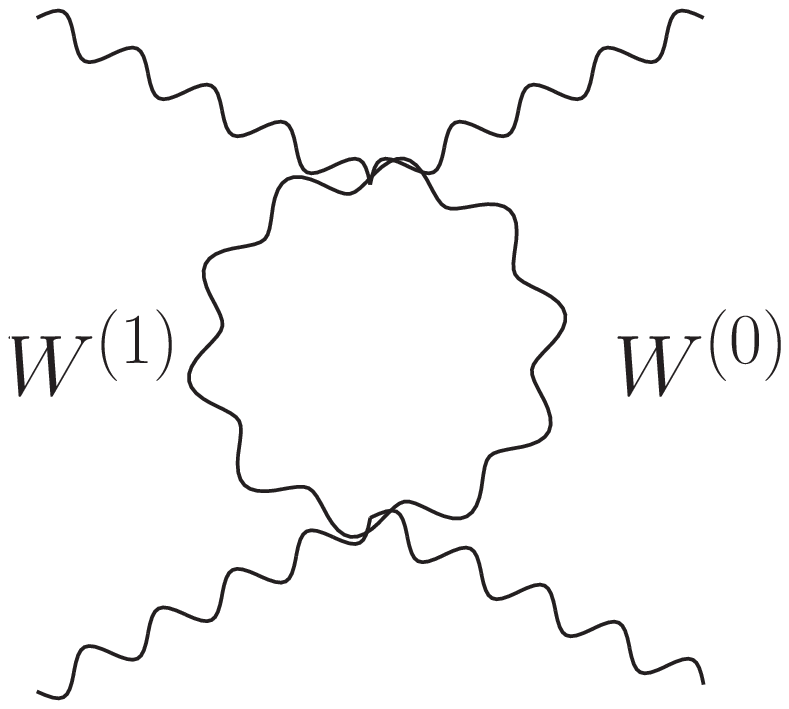} \\							
\vspace{5mm}
\includegraphics[width=.17\textwidth]{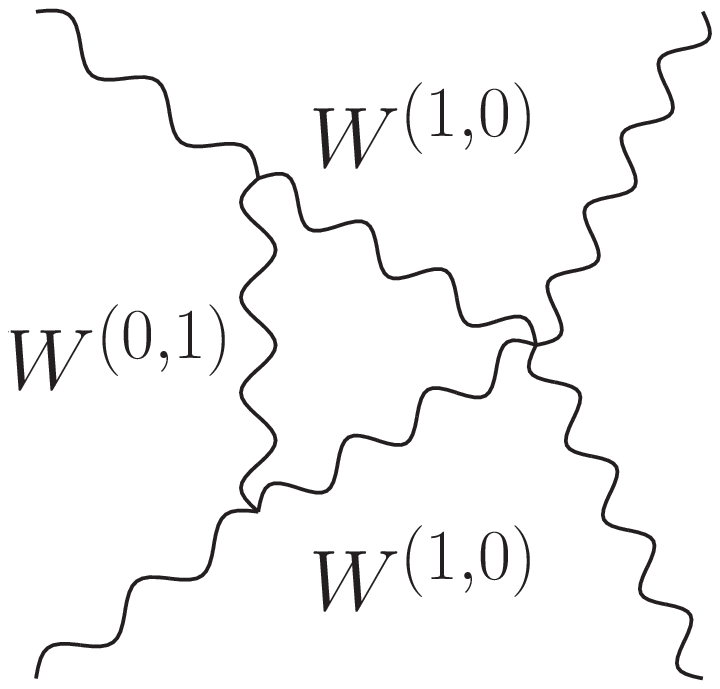} \hspace{.02\textwidth}	
\includegraphics[width=.17\textwidth]{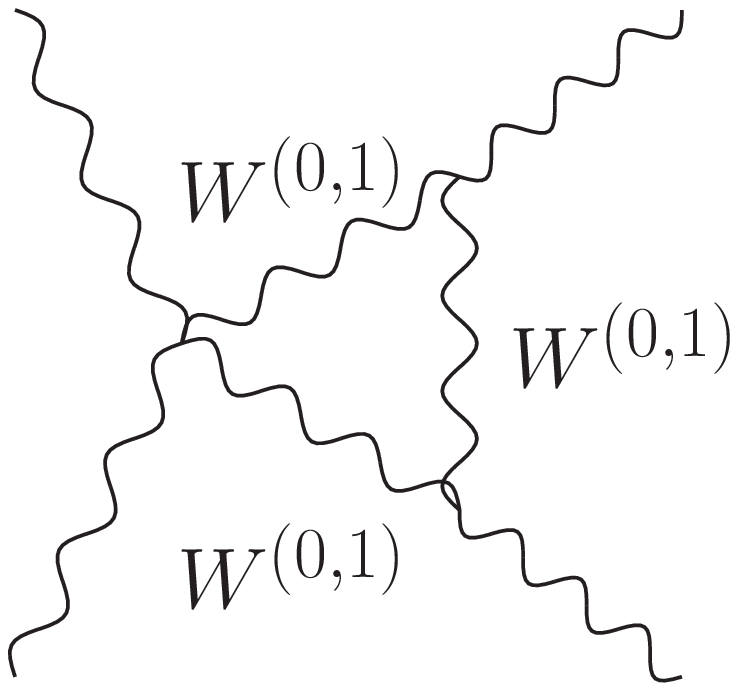} \hspace{.02\textwidth}	
\includegraphics[width=.17\textwidth]{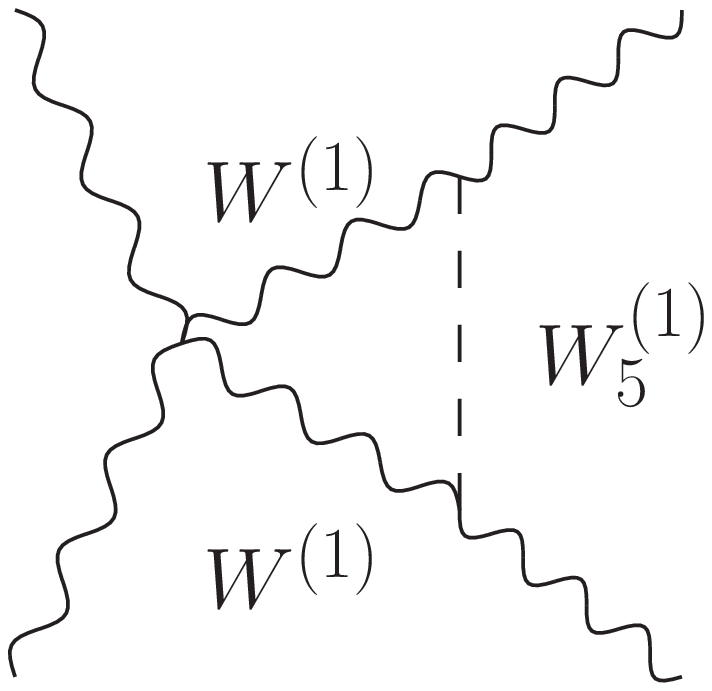} \hspace{.02\textwidth}
\includegraphics[width=.17\textwidth]{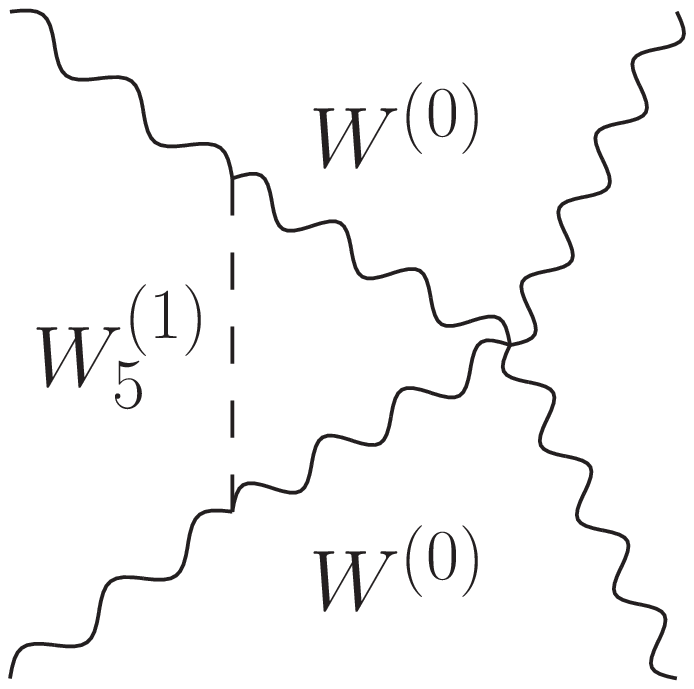} \hspace{.02\textwidth}		
\includegraphics[width=.17\textwidth]{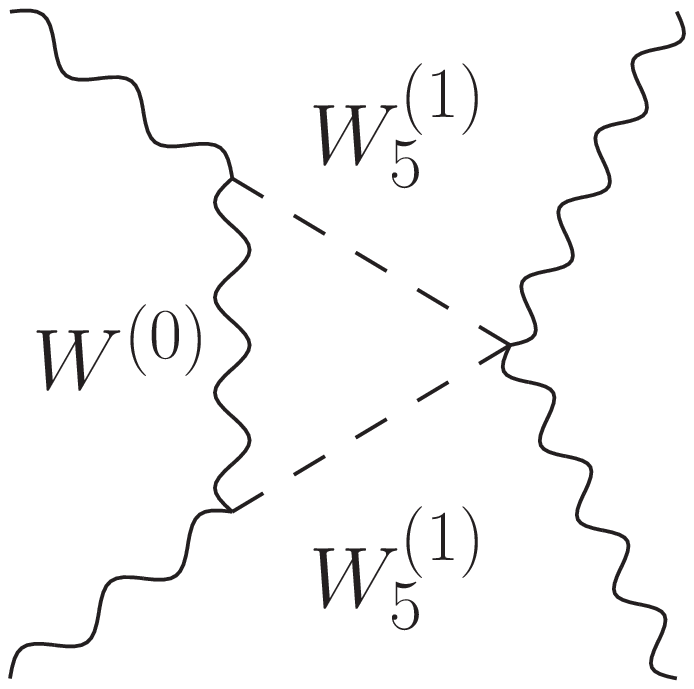} \\							
\vspace{5mm}
\includegraphics[width=.17\textwidth]{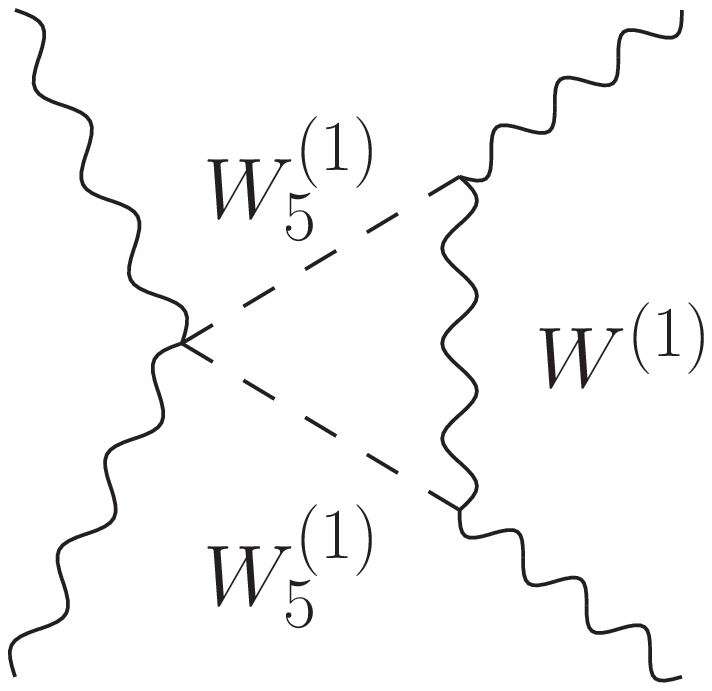} \hspace{.02\textwidth}		
\includegraphics[width=.17\textwidth]{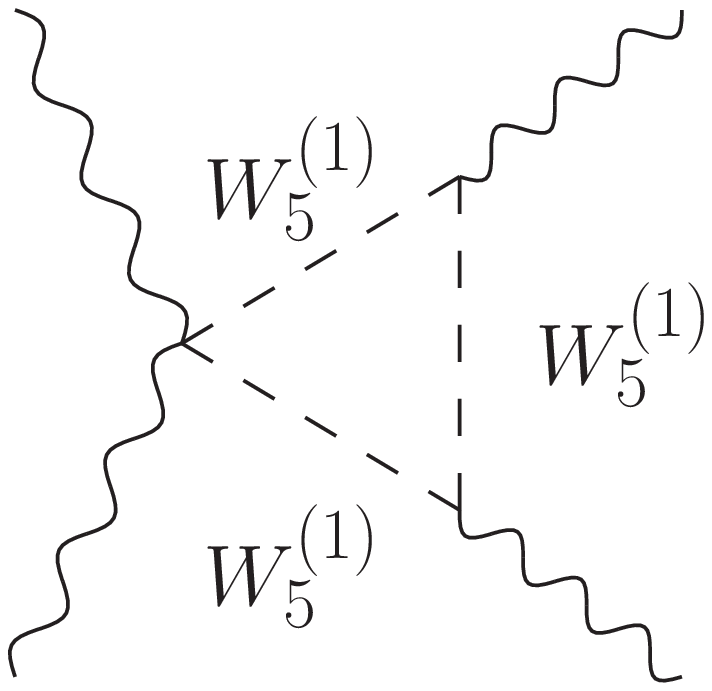} \hspace{.02\textwidth}		
\includegraphics[width=.17\textwidth]{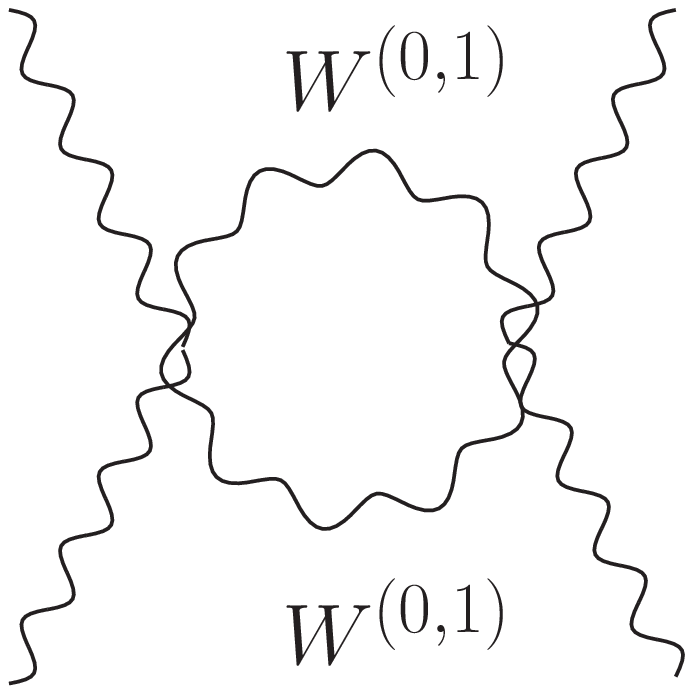} \hspace{.02\textwidth}	
\includegraphics[width=.17\textwidth]{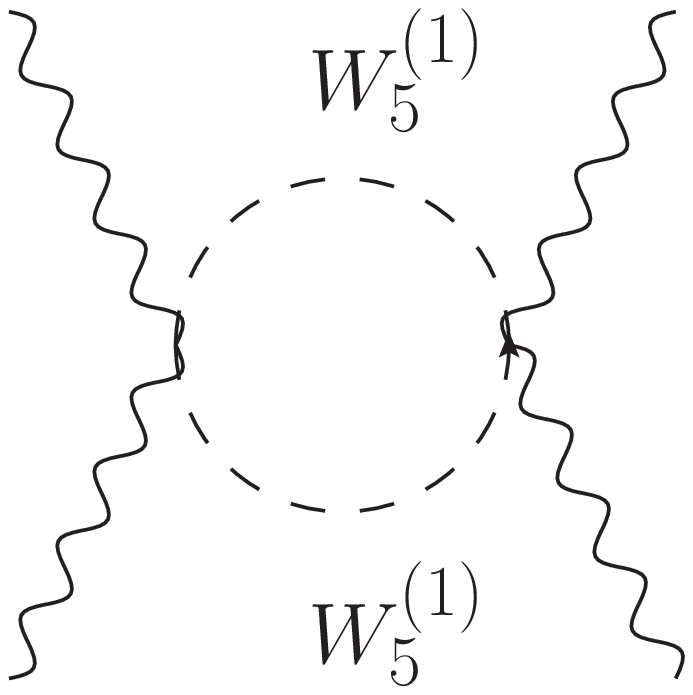}	
\caption{The one-loop level Feynman diagrams involving internal gauge fields for the process $Z^1 Z^1 \to \gamma \gamma$. For each diagram in the first four rows, there is also a corresponding one with crossed final states, making the total number equal to 60. The notation $A^{(0,1)}$ simply indicates that there are two different choices for the flow of KK number in the diagram, and the first or second choice should be used throughout the diagram.} \label{fig:FeynmanDiagramsGauge}
\end{figure}

\begin{figure}[htb]
\includegraphics[width=.17\textwidth]{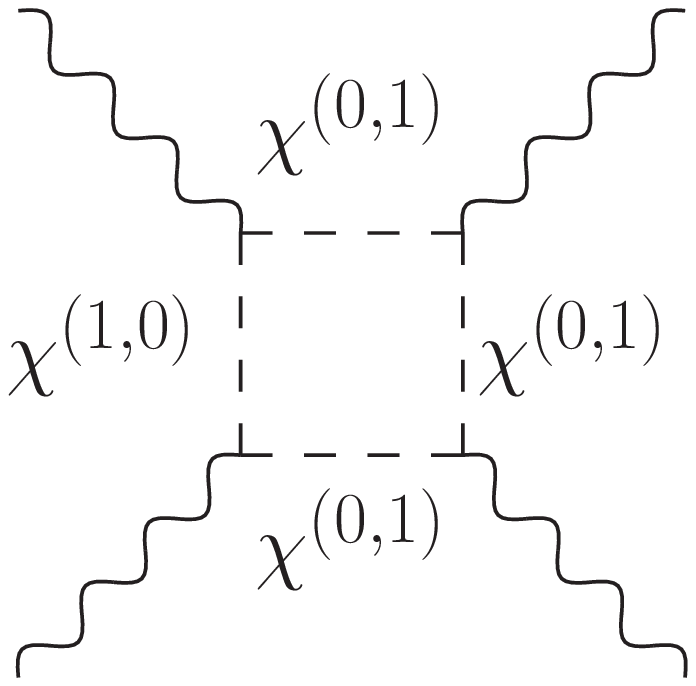} \hspace{.02\textwidth}
\includegraphics[width=.17\textwidth]{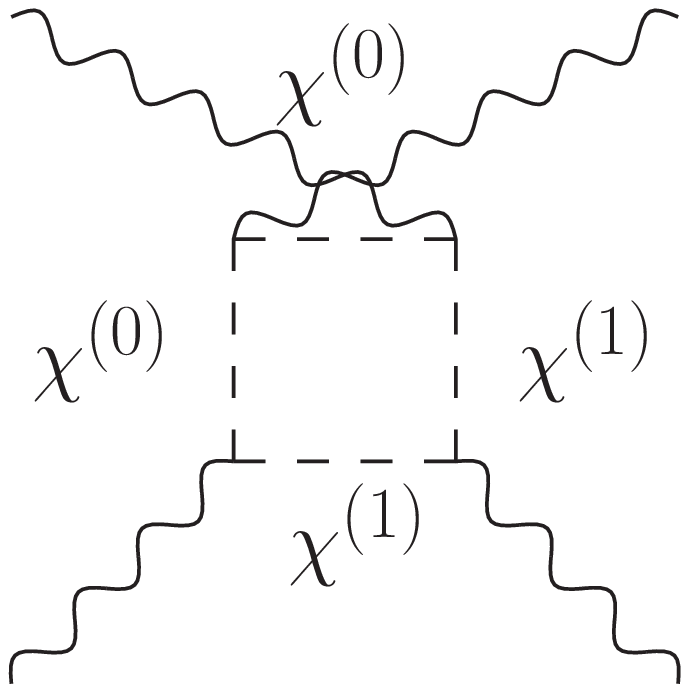} \hspace{.02\textwidth}
\includegraphics[width=.17\textwidth]{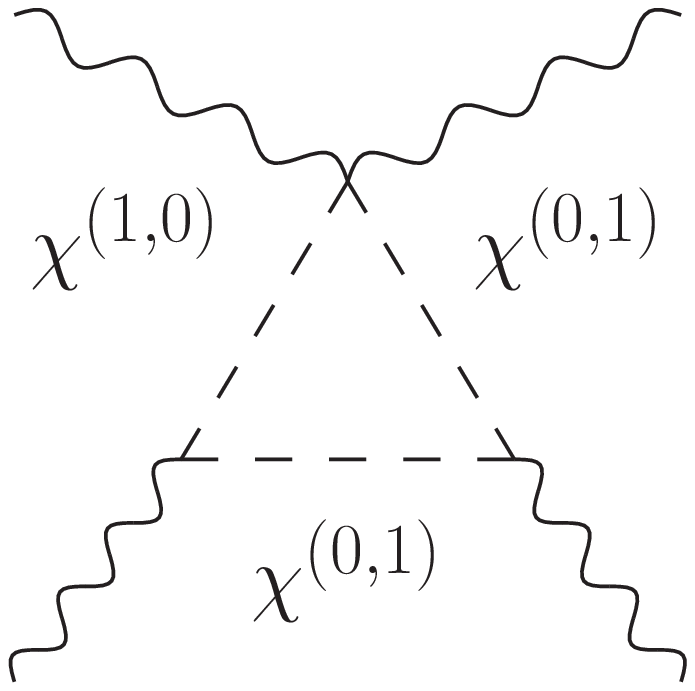} \hspace{.02\textwidth}
\includegraphics[width=.17\textwidth]{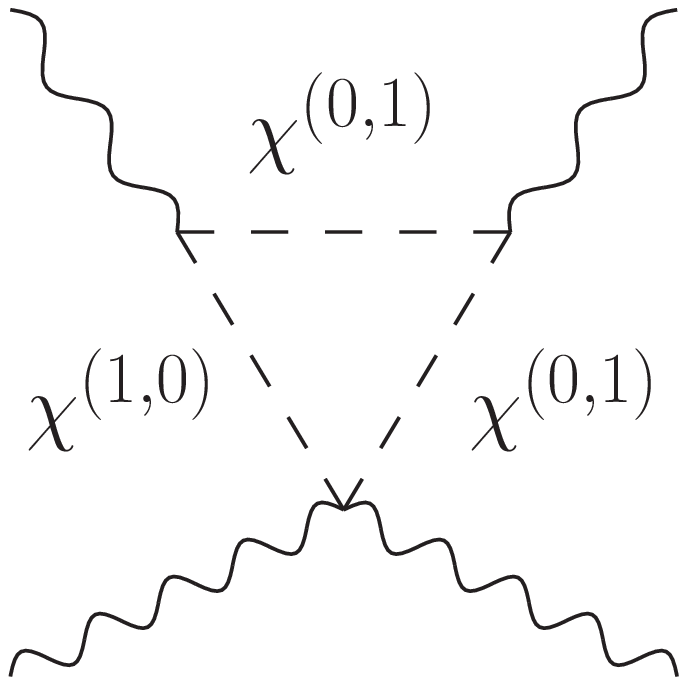} \hspace{.02\textwidth}
\includegraphics[width=.17\textwidth]{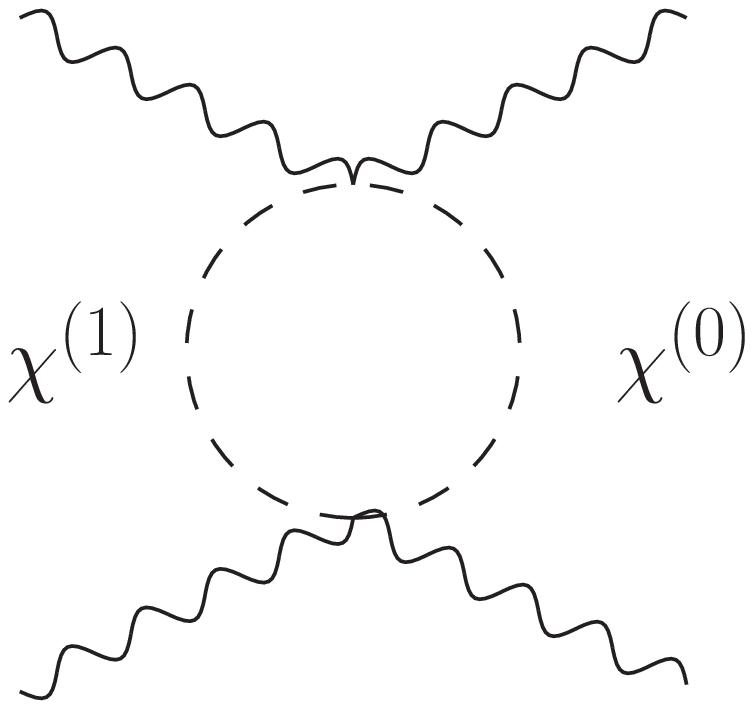} \\
\vspace{5mm}
\includegraphics[width=.17\textwidth]{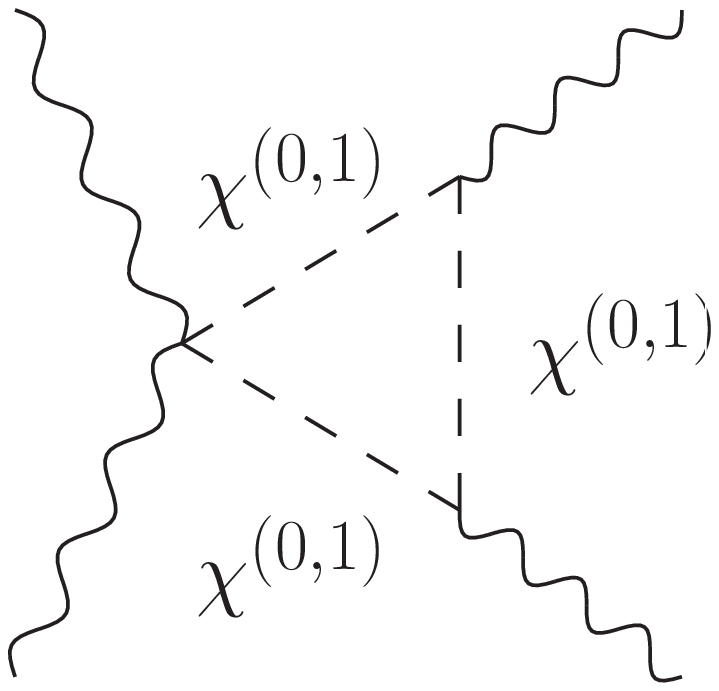} \hspace{.02\textwidth}
\includegraphics[width=.17\textwidth]{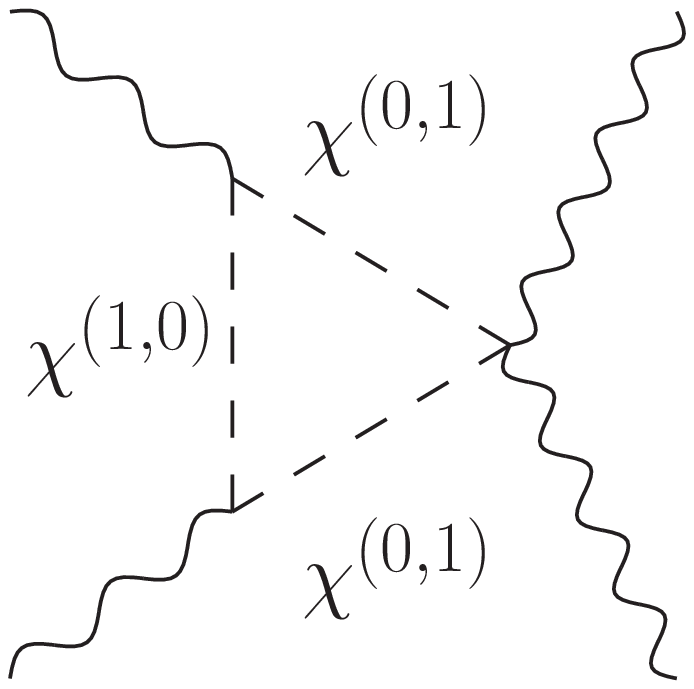} \hspace{.02\textwidth}
\includegraphics[width=.17\textwidth]{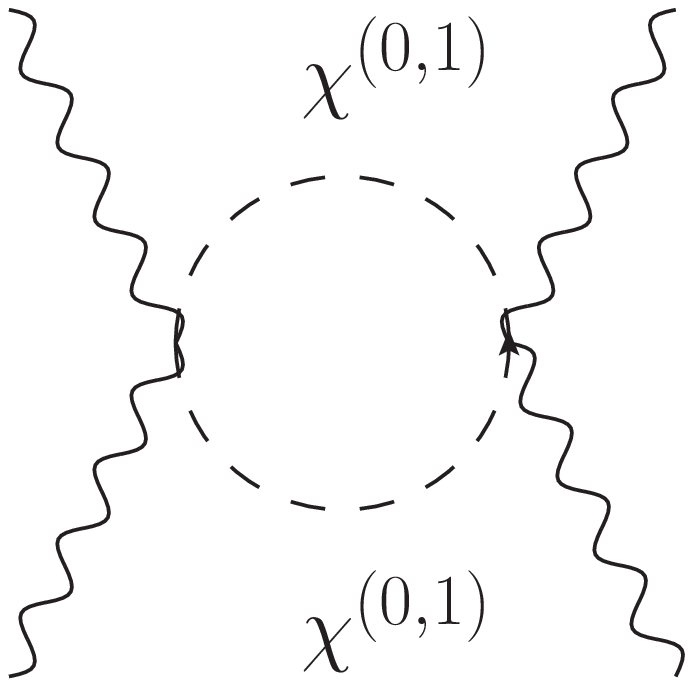} \hspace{.02\textwidth}
\caption{The one-loop level Feynman diagrams involving internal Higgs scalars for the process $Z^1 Z^1 \to \gamma \gamma$. Here, $\chi$ denotes the charged component of the Higgs doublet. For each diagram in the first row, there is also a corresponding one with crossed final states, making the total number equal to 22. The notation $A^{(0,1)}$ simply indicates that there are two different choices for the flow of KK number in the diagram, and the first or second choice should be used throughout the diagram.} \label{fig:FeynmanDiagramsHiggs}
\end{figure}

We have identified a total number of 88 Feynman diagrams that contribute to the process $Z^1 Z^1 \to \gamma \gamma$ at one-loop level. In Figs.~\ref{fig:FeynmanDiagramsFermions}, \ref{fig:FeynmanDiagramsGauge}, and \ref{fig:FeynmanDiagramsHiggs}, we show the diagrams with internal fermions, gauge fields, and Higgs scalars, respectively.

\section{Feynman rules}\label{sec:FeynmanRules}

In this appendix, we present a number of Feynman rules for the UED model that are, to our knowledge, previously not existing in the literature. The fields $W_5^\pm$ denote linear combinations of the charged ${\rm SU}(2)_{\rm L}$ gauge bosons that are analogous to their four-dimensional counterparts. The ghost fields $c^\pm$ are defined as $c^\pm = (c^1 \mp c^2)/\sqrt{2}$, \ie, they are not each other's Hermitian conjugates. The constant $g$ is the ${\rm SU}(2)_{\rm L}$ gauge coupling constant, $e$ is the electromagnetic coupling constant, and $M_1 = R^{-1}$ is the mass scale of the first excited KK level. For the Higgs doublet, we use the convention
\begin{equation}
	\Phi = \left( \begin{array}{cc} {\rm i} \chi^+ \\ \frac{1}{\sqrt{2}} (h + {\rm i} \chi^0) \end{array} \right).
\end{equation}

\subsection*{Gauge boson self-interactions}

\begin{align*}
\parbox{45mm}{\includegraphics[width=45mm]{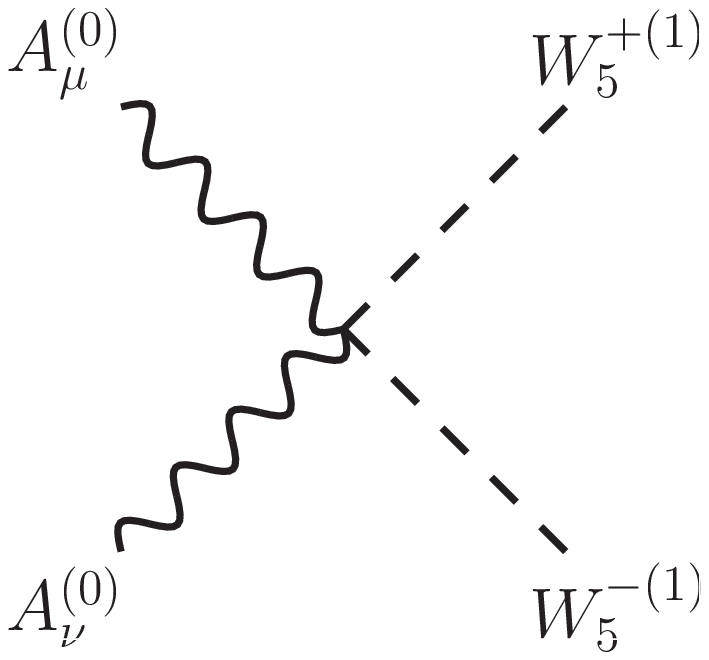}} & \hspace{0.5cm}= 2{\rm i} e^2 g^{\mu\nu} &
\parbox{45mm}{\includegraphics[width=45mm]{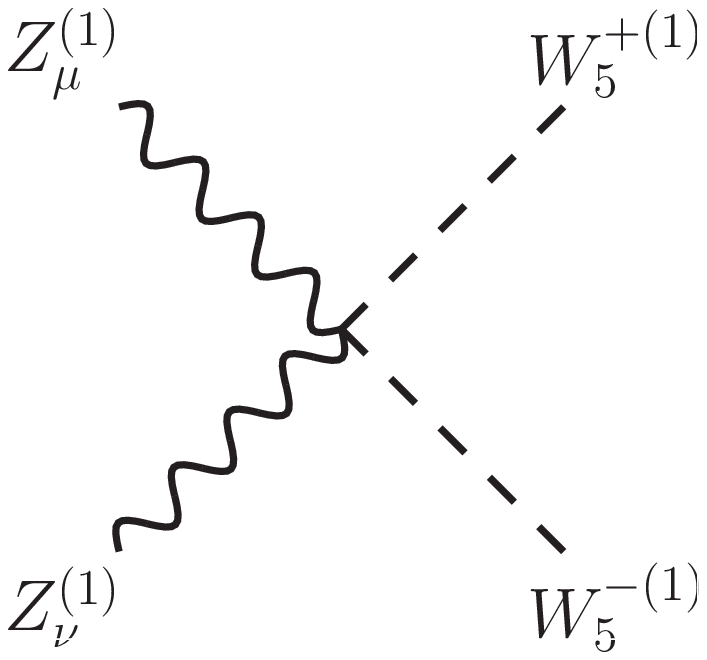}} & \hspace{0.5cm}= {\rm i} g^2 g^{\mu\nu} \\
\\
\parbox{45mm}{\includegraphics[width=45mm]{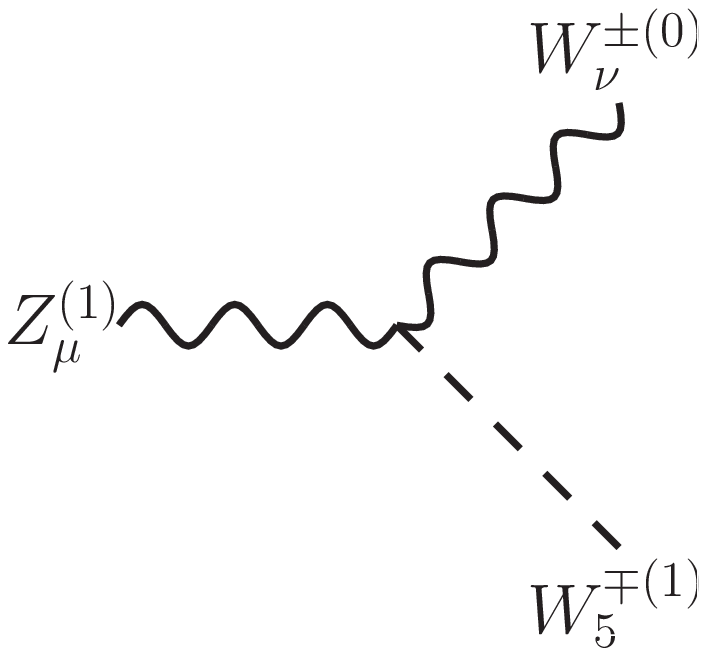}} & \hspace{0.5cm}= \pm g M_1 g^{\mu\nu} &
\parbox{45mm}{\includegraphics[width=45mm]{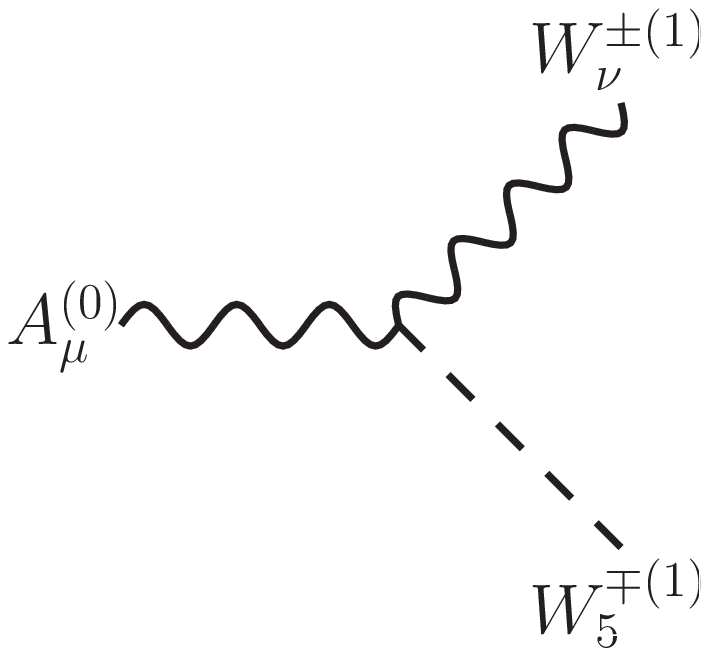}} & \hspace{0.5cm}= \mp e M_1 g^{\mu\nu} \\
\\
\parbox{45mm}{\includegraphics[width=45mm]{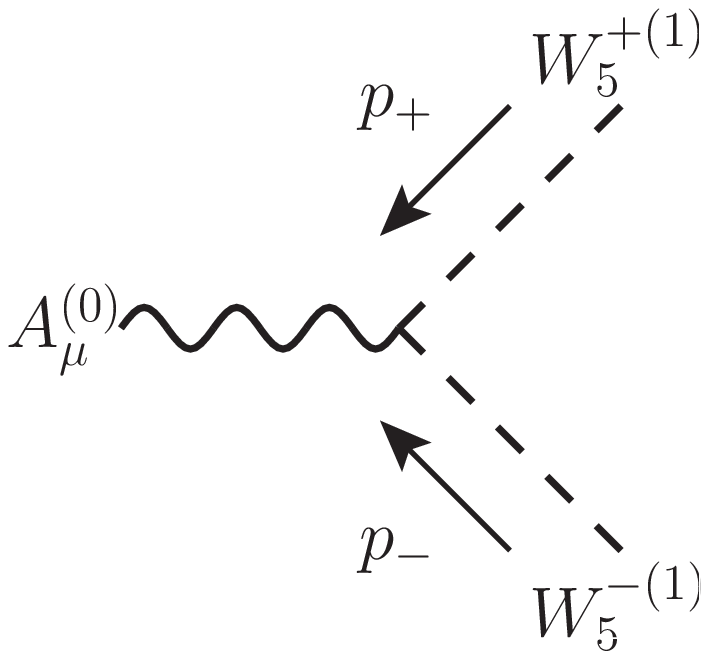}} & \hspace{0.5cm}= - {\rm i} e (p_+ - p_-)^{\mu}
\end{align*}

\subsection*{Ghost interactions with gauge fields}

\begin{align*}
\parbox{45mm}{\includegraphics[width=45mm]{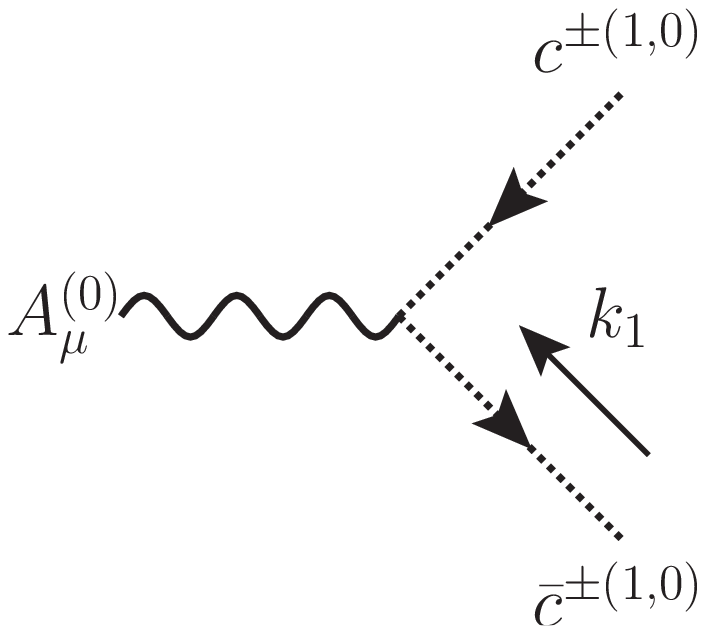}} & \hspace{0.5cm}= \pm e k_1^{\mu} &
\parbox{45mm}{\includegraphics[width=45mm]{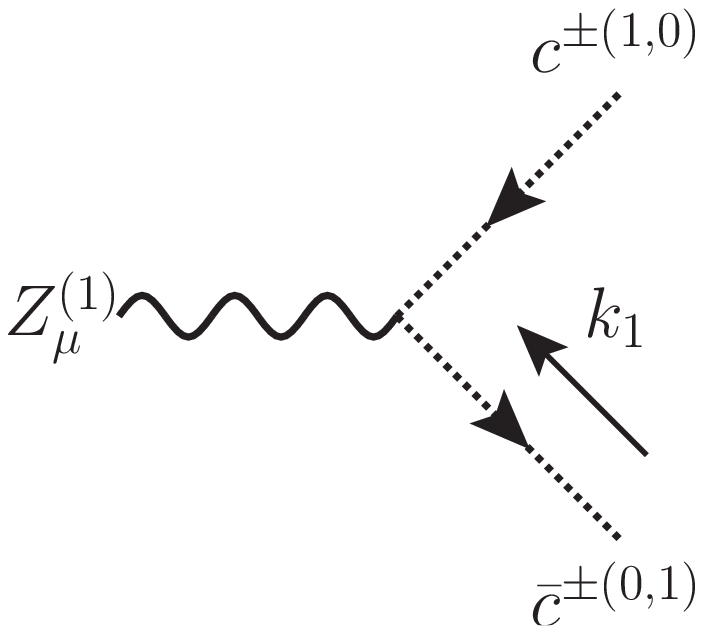}} & \hspace{0.5cm}= \pm g k_1^{\mu}
\end{align*}

\subsection*{Gauge boson-Higgs scalar interactions}

\begin{align*}
\parbox{45mm}{\includegraphics[width=45mm]{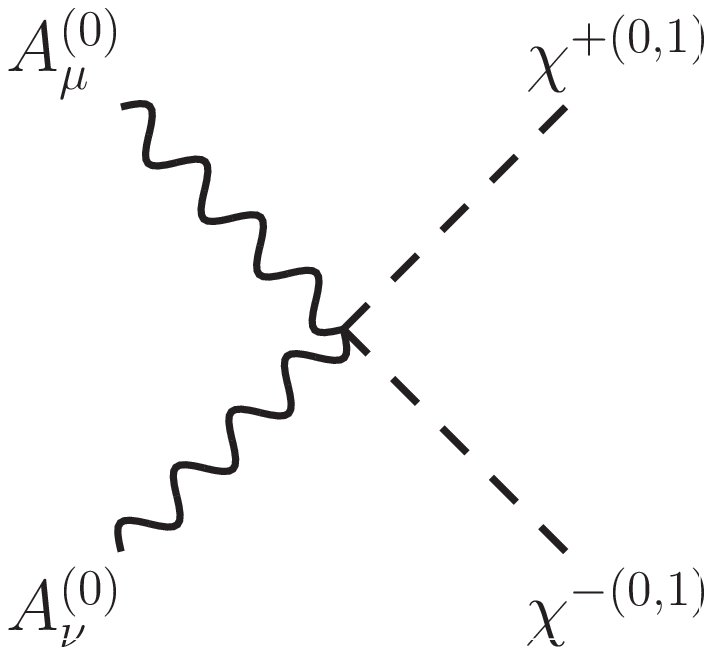}} & \hspace{0.5cm}= 2{\rm i} e^2 g^{\mu\nu} &
\parbox{45mm}{\includegraphics[width=45mm]{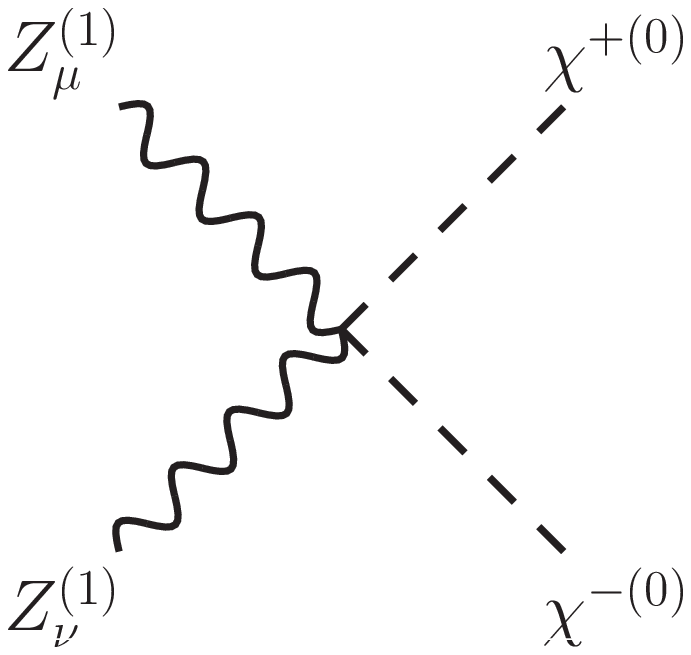}} & \hspace{0.5cm}= {\rm i} \frac{g^2}{2} g^{\mu\nu} \\
\\
\parbox{45mm}{\includegraphics[width=45mm]{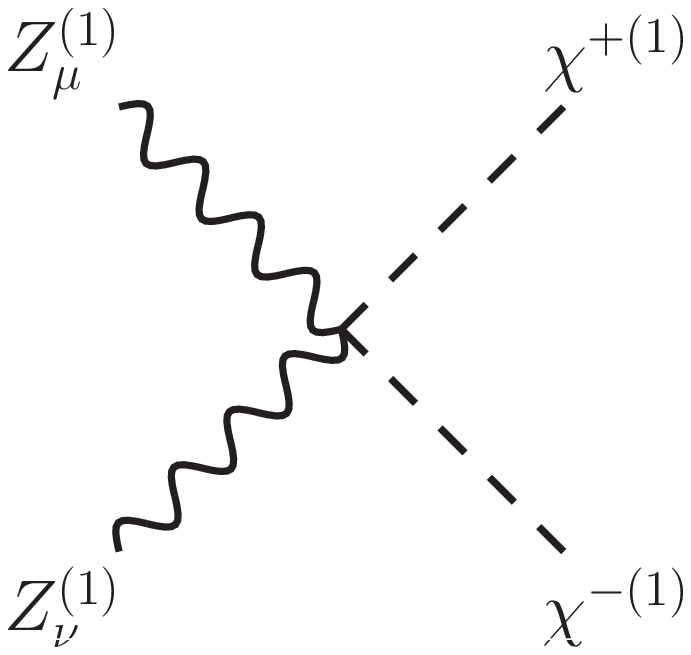}} & \hspace{0.5cm}= {\rm i} \frac{3 g^2}{4} g^{\mu\nu} &
\parbox{45mm}{\includegraphics[width=45mm]{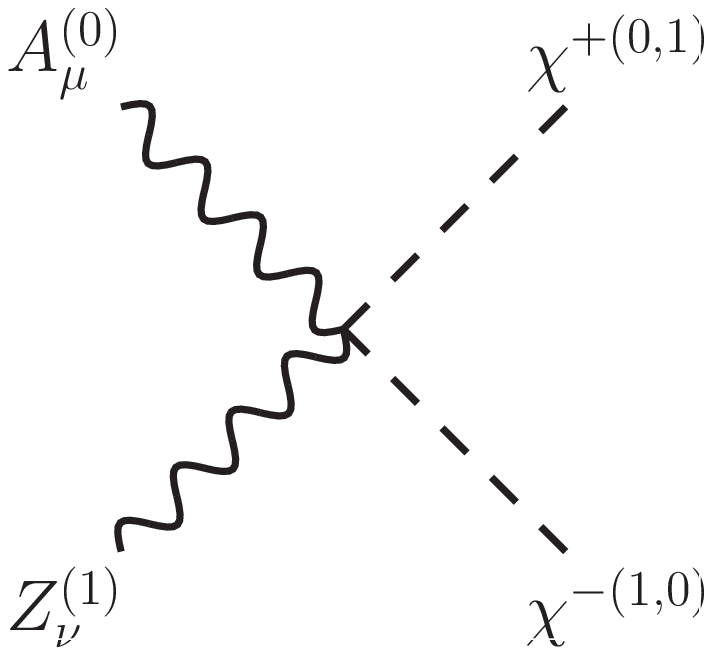}} & \hspace{0.5cm}= {\rm i} e g g^{\mu\nu} \\
\\
\parbox{45mm}{\includegraphics[width=45mm]{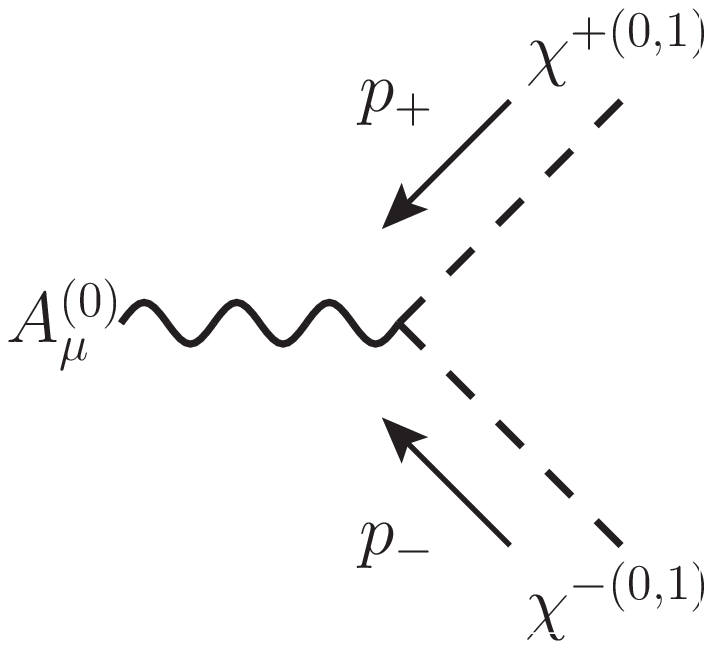}} & \hspace{0cm}= \pm {\rm i} e (p_+ - p_-)^{\mu} &
\parbox{45mm}{\includegraphics[width=45mm]{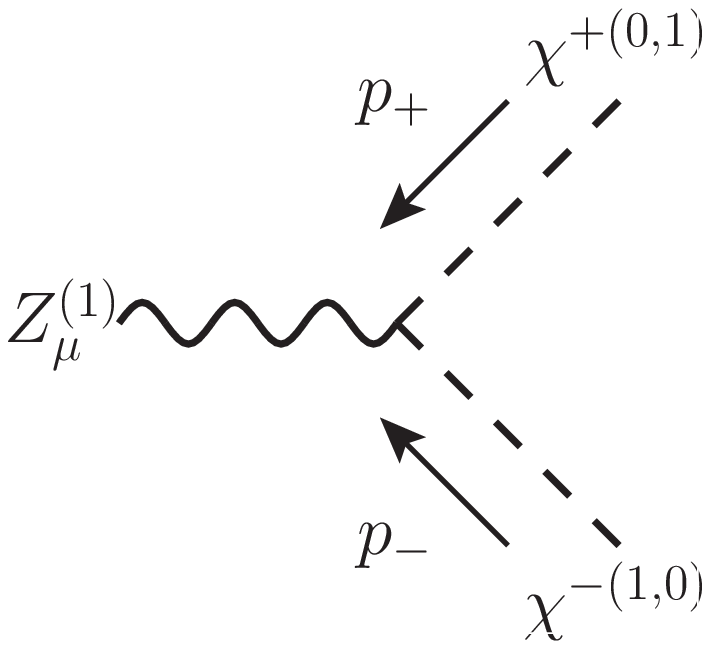}} & \hspace{0cm}= \pm {\rm i} \frac{g}{2} (p_+ - p_-)^{\mu}
\end{align*}

\section{Cancellation of divergences} \label{sec:Divergences}

In this appendix, we argue why potential divergences in our calculation can be ignored. In general, using dimensional regularization, a one-loop amplitude can be written in the form
\begin{equation}
	\mathcal{M}_\text{1-loop} = \sum_i \left( \frac{\Delta_i}{\epsilon} + F_i \right),
	\label{eq:1-loop-amp_DivFin}
\end{equation}
where $\Delta_i/\epsilon$ are divergences and $F_i$ are the finite parts. Per definition, a theory is (perturbatively) renormalizable at the one-loop level if all divergences appearing in the one-loop diagrams can be absorbed in counter terms, which must exist as explicit tree-level vertices in the Lagrangian. Hence, for any process that lacks such a tree-level vertex (such as the annihilation $Z^1 Z^1 \to \gamma \gamma$ that is investigated in this paper) the coefficients $\Delta_i$ must sum to zero. 

Now, if we consider only a finite subset of diagrams, e.g., taking into account only the contributions from the lowest KK-modes, we cannot expect that the divergent parts will cancel. Although, within this approximation, the interactions are given by a finite number of operators which would seem to be renormalizable by dimensional analysis, a complete gauge theory is generally not obtained. This means that the cancellation mechanisms arising from gauge invariance, \ie, the Ward identity, do not work out. On the other hand, if we sum all diagrams at a certain order in the coupling constant, for the complete infinite KK-tower, we will recover the full five-dimensional theory at this order in the coupling. However, the full five-dimensional theory is non-renormalizable~\cite{Cheng:2002iz}, since all coupling constants in the five-dimensional Lagrangian have negative mass dimensions. Hence, there is no reason to expect that $\sum_i \Delta_i=0$ in this case either.

However, this issue is only a reflection of the fact that the UED model is just an effective low-energy theory, which must be replaced by a UV-complete theory at some high-energy scale. The additional contributions to the amplitude in Eq.~\eqref{eq:1-loop-amp_DivFin} from the UV part of the theory lead to the full 1-loop amplitude
\begin{equation}
	\mathcal{M}_\textrm{1-loop}^{\rm total} = \mathcal{M}_\textrm{1-loop} + \frac{\Delta_{\rm UV}}{\epsilon} + F_{\rm UV}.
	\label{eq:1-loop-amp_tot}
\end{equation}
If the UV-complete theory is renormalizable, then it must hold that
\begin{equation}
	\sum_i \Delta_i + \Delta_{\rm UV}=0.
	\label{eq:UV_renorm}
\end{equation}
Without detailed knowledge of the UV completion, it is not possible to show this cancellation explicitly, even in principle. However, consistency of the extra-dimensional framework at high-energy scales requires that it does exist and that it is renormalizable, which means that the divergences $\Delta_i$ arising from the low-energy part do not spoil the predictions. Although there exist examples of such renormalizable UV-completions~\cite{ArkaniHamed:2001ca}, the exact form of this underlying theory is irrelevant for our purposes. 

The size of the contribution from the UV sector, $F_{\rm UV}$, can be estimated to be of the order of $M_{\rm KK} / \Lambda_{\rm UV} = R^{-1} / \Lambda_{\rm UV}$, where $\Lambda_{\rm UV}$ is the energy scale of the UV sector, i.e., the cutoff scale of the effective extra-dimensional model. In the five-dimensional UED model, the cutoff scale is of the order of $20 R^{-1}$ \cite{SekharChivukula:2001hz}, i.e., $R^{-1} / \Lambda_{\rm UV} \sim 5 \cdot 10^{-2}$. In the worst-case scenario, when the UV sector is strongly coupled, this number could be multiplied by a factor which is as large as $4\pi$, in which case $F_{\rm UV} / \mathcal{M}_\textrm{1-loop} = \mathcal{O} (1)$.

As for the contributions from higher KK-modes, there are diagrams that conserve KK number and diagrams that break conservation of KK number, but preserve KK parity. The vertices that break KK number are already loop-suppressed, and hence, the corresponding diagrams are expected to give only a small contribution. The diagrams that do conserve KK number necessarily contain a larger number of massive propagators than the diagrams that we take into account (all of which contain massless propagators), and therefore, these diagrams are also suppressed, due to the minimum size of $1/R$. Quantitatively, the effect of higher KK-modes was numerically investigated in Ref.~\cite{Bergstrom:2004nr} for the case of internal fermions in the $B^1 B^1 \to \gamma \gamma$ process. In that work, it was found that the contribution is only a few percent, and we expect a similar result for our case.

To summarize the effects of the finite contributions that we neglect, the contribution from the higher KK-modes is negligible, while the UV sector could possibly give a contribution as large as $\mathcal{O} (1)$, if it is strongly coupled. In that case, our results are valid only up to an order of magnitude. On the other hand, if the UV sector is weakly coupled, its contributions are suppressed by powers of small coupling constants, and then, our results are more accurate.

Finally, any loop-calculation involving gauge theories in UEDs where a cancellation of divergences does take place at each KK level has the feature to resemble some renormalizable theory. For the process $B^1 B^1 \to \gamma \gamma$, the Ward identity and the cancellation of divergences do hold at each KK level, as was confirmed in Ref.~\cite{Bergstrom:2004nr}. This is a natural consequence of the fact that this process, at any KK level, is essentially a copy of the corresponding process in ordinary Quantum Electrodynamics (QED). Hence, the cancellation of divergences in QED simply translates to the UED process level-by-level. The same statement is true for the fermionic part of the process $Z^1 Z^1 \to \gamma \gamma$, which our calculations confirm, but not for the full process.

\end{document}